\documentclass{jfm}
\usepackage{graphicx}
\usepackage{epstopdf, epsfig}

\usepackage{graphicx}
\usepackage{svg}
\usepackage{subcaption}
\usepackage{dcolumn}
\usepackage{bm}

\usepackage[utf8]{inputenc}
\usepackage[T1]{fontenc}
\usepackage{mathptmx}
\usepackage{etoolbox}
\usepackage{amsmath}
\usepackage{amssymb}

\usepackage{etoolbox}
\usepackage{tikz}
\usepackage{comment}

\newcommand{\sqdiamond}[1][fill=black]{\tikz [x=1.2ex,y=1.85ex,line width=.1ex,line join=round, yshift=-0.285ex] \draw  [#1]  (0,.5) -- (.5,1) -- (1,.5) -- (.5,0) -- (0,.5) -- cycle;}%
\newcommand{\MyDiamond}[1][fill=black]{\mathop{\raisebox{-0.275ex}{$\sqdiamond[#1]$}}}

\newrobustcmd*{\mysquare}[1]{\tikz{\filldraw[draw=#1,fill=#1] (0,0)
rectangle (0.2cm,0.2cm);}}

\newrobustcmd*{\mycircle}[1]{\tikz{\filldraw[draw=#1,fill=#1] (0,0) circle [radius=0.1cm];}}

\shorttitle{Engulfment of a drop on solids coated by thin and thick fluid films}

\shortauthor{C. Zhao, V.R. Kern, A. Carlson and T. Lee}

\title{Engulfment of a drop on solids coated by thin and thick fluid films}

\author{Chunheng Zhao\aff{1}, Vanessa R. Kern\aff{2}, 
  Andreas Carlson\corresp{\email{acarlson@math.uio.no}}\aff{2}
  \and Taehun Lee\aff{1}}

\affiliation{\aff{1}Department of Mechanical Engineering, City College of New York, New York, NY 10031, USA.
\aff{2}Department of Mathematics, Mechanics Division, University of Oslo, Oslo 0316, Norway}

\begin{document}

\maketitle

\begin{abstract}
When an aqueous drop contacts an immiscible oil film, it displays complex interfacial dynamics. Upon contact the oil spreads onto the drop’s liquid-air interface, first forming a curvature that drives an apparent drop spreading motion and later fully engulfing the drop. We study this flow using both 3-phase Lattice-Boltzmann simulations based on the conservative phase field model and experiments. Inertially and viscously limited dynamics are explored using the Ohnesorge number $Oh$ as a function of $R/H$, the ratio between the initial drop radius $R$ and the film height $H$. Both regimes show that the apparent spreading radius $r$ is fairly independent of the film height, and scales with time $T$ as $r\sim T^{1/2}$ for $Oh\ll 1$ and for $Oh\gg 1$ as $r\sim T^{2/5}$. For $Oh\gg 1$ we show experimentally that this immiscible apparent spreading motion is analogous with the miscible drop-film coalescence case. Contrary to the apparent spreading, however, we find that the late time engulfment dynamics and final interface profiles are significantly affected by the ratio of $H/R$.
\end{abstract}


\section{\label{sec:level1}Introduction}    

Over the last decade, pre-wetted surfaces and slippery liquid-infused porous substrates (SLIPS) have attracted significant academic and industrial attention due to their anti-icing, low friction and multifunctional properties  \citep{wong2011bioinspired,anand2012enhanced,smith2013droplet,subramanyam2013ice,rykaczewski2013mechanism,solomon2014drag,hao2015superhydrophobic,park2016condensation,villegas2019liquid}. 
These surfaces rely on a stable viscous fluid film to separate drops and other particles from the supporting solid, offering a relatively slippery, pressure-stable surface compared to their solid counterparts \citep{solomon2016lubricant,huang2019fabrications,bohn2004insect}. 
Wetting on soft materials has over the last few years emerged as a topical field in fluid mechanics, where pronounced deformations of the soft substrate near the contact line greatly affect the spreading dynamics \citep{andreotti2020statics}. A drop contacting a substrate coated with an immiscible fluid is one type of soft wetting problem that 
has received far less attention \citep{smith2013droplet,daniel2017oleoplaning,daniel2018origins}.

When a drop of an immiscible fluid comes in contact with a SLIPS surface, it first displays a spreading like behaviour \citep{carlson2013short} with the flow driven by capillarity and limited by viscous friction. Here, the spreading dynamics depend on the ratios between the material properties of the fluid phases i.e. viscosity $\eta$, surface tension $\sigma$, density $\rho$, as well as the ratio between the drop's radius $R$ and the film's thickness $H$. \cite{cuttle2021engulfment} decomposed the interfacial dynamics generated when an aqueous drop contacts an oil layer into different stages, first short-time spreading, then the complete engulfment of the drop by the oil film, followed by the drop being "pushed" into the bath/film. 
Depending on the spreading factor, finite contact angles can form between the drop and the film, creating a liquid lens \citep{de2004capillarity,kuiper2004variable,cheng2007dielectrically,hack2020self}. Contrary to capillary spreading on a solid substrate \citep{bonn2009wetting}, the fluid film introduced by the SLIPS alleviates the stress singularity at the contact line while generating additional viscous dissipation that extends into the fluid film. 
A detailed description of the dynamics of the contact line motion and the formation of a film on the drop for this soft-wetting phenomenon is yet to be described, as it is experimentally challenging to measure the interfacial dynamics at these small time and length scales that are additionally complicated by the presence of three fluid phases. In this article we describe the details of the small scale interfacial dynamics of the three phase flow by using the Lattice Boltzmann method (LBM) to solve the conservative phase field model, with simulation results compared directly to experiments. In addition, we address how the thickness of the coating film affects both the inertially limited and viscously limited drop engulfment dynamics, and describe where the dissipation is localized in these flows.

During the apparent spreading post drop-film contact, the characteristic velocity of the contact line is set by the capillary velocity $\sigma/\eta$. Thus, it is natural to describe the drop's dynamics using the Ohnesorge number, $Oh=\eta_o/\sqrt{\rho_o \sigma_{wo}R}$, where the subscript $o$ represents the oil film and the subscript $w$ represents the water drop. 
For thin films $ H/R \ll 1$, the spreading dynamics have been shown to scale with an inertial time scale $t_\rho=\sqrt{\rho R^3/\sigma_{ao}}$ when $Oh\ll1$ and with a viscous time scale $t_\eta=\eta_o R/\sigma_{ao}$ for $Oh\gg1$, where the subscript $a$ represents air. 
The evolution of the drop's apparent spreading radius can then be found to follow the scaling law $r/R\sim (T/t_\rho)^{1/2}$ when $Oh\ll1$ and $r/R\sim (T/t_\eta)^{7/20}$ when $Oh\gg1$ \citep{carlson2013short}.
For thick films $H/R\gg1$, \cite{cuttle2021engulfment} described the long time entrapment and transport into the bath. 
Despite these recent considerations, however, drop engulfment dynamics for moderate $H/R$ have been sparsely considered and are the focus of this work.

When an aqueous drop contacts an immiscible liquid film, the resulting interfacial dynamics can be interpreted as a combination of two widely studied phenomena, capillary driven contact line motion  \citep{de1985wetting,de2004capillarity,bonn2009wetting,snoeijer2013moving} and miscible drop coalescence \citep{orme1997experiments,frising2006liquid,yao2005coalescence}. Both of these phenomena have been characterised in both the inertial and viscous regimes \citep{guido1998binary,duchemin2003inviscid,lee2006eliminating,paulsen2011viscous,baroudi2015dynamics,luo2016effect}, as well as in more complex situations such as under electric fields  \citep{mugele2005electrowetting} and with complex surface properties \citep{ kapur2007morphology,mazutis2012selective,aarts2005hydrodynamics,burton2007role}. 

During rigid solid surface spreading, the radius of the drop's contact line has been used as a proxy to characterise the spreading dynamics. For low viscosity drops experiencing inertially limited contact-line motion, the spreading radius $r$ has been predicted to follow a power-law relation as $r/R\sim (T/t_{c})^{1/2}$, where $T$ is the time and $t_{c}$ the characteristic time scale \citep{carlson2011dissipation,eddi2013short,carlson2013short,biance2004first,courbin2009dynamics}. 
For viscously-limited spreading on perfectly wetting substrates, analogies to viscous coalescence have been made, motivated by the curvature between the drop and its pre-cursor film \citep{eddi2013short,bird2008short}.
In non-perfectly wetting systems, the spreading radius $r$ has been found to follow a power-law-like behaviour $r \sim t^\beta$, where $\beta$ has been shown to vary with the surface wettability at short time \citep{winkels2012initial,mitra2016understanding, bird2008short}, before transitioning to a tanner's law like spreading behaviour at long time \citep{tanner1979spreading}. These differences in the power-law behaviour in the short time have been shown to be a function of contact-line friction, which itself is a function of surface wettability, and recent scalings have been shown to collapse these differences \citep{carlson2012contact}.

During drop-drop coalescence, much focus has been placed on the shape and evolution of the neck radius $r$ that forms as the two miscible drops touch, as it gives a way to characterise the coalescence dynamics and to identify the different flow regimes \citep{eggers1999coalescence,aarts2005hydrodynamics}. 
In inertially limited systems, the neck radius $r$ expands rapidly post coalescence and the curvature, $\kappa \sim R/r^2$, generates a capillary pressure $p_\sigma\sim\sigma \kappa\sim \sigma R/r^2$. Tracing the stream lines from the drop interior to the interface, it follows then from the Bernoulli equation that $p_\sigma=\frac{\rho}{2}u_r^2 $, where $u_r=\textrm{d}r/\textrm{d}t$ is the neck velocity. Consequently the evolution of spreading radius in time can be written as $r/R\sim(T/t_\rho)^{1/2}$ \citep{eggers1999coalescence}, where $t_\rho=\sqrt{\rho R^3/\sigma_{ao}}$ is the inertial time scale. In viscously limited systems, the momentum equation simplifies to the Stokes equation $ \eta\nabla^2u=\nabla p_\sigma $ and a viscous time scale defines the dynamics $t_{\eta}=\eta R / \sigma_{ao}$. Here the evolution of spreading radius in time can be written as $ r/R\sim(T/t_\eta)^{\beta} $, where $\beta$ is determined by the viscous dissipation in the liquid. Simulations and theoretical analysis of two equal-sized drops coalescing indicates $\beta=1/2\sim 1$, with potential for logarithmic correction at early times \citep{eggers1999coalescence,hopper1984coalescence,paulsen2014coalescence}. 

In recent years, simulations of contact line spreading and drop coalescence have helped to reveal additional dynamics in both the viscous and inertial regimes that are otherwise challenging or nearly impossible to observe experimentally \citep{lee2006eliminating,paulsen2012inexorable, baroudi2015dynamics,xia2019universality, baroudi2020effect,anthony2020initial,anthony2017scaling}. 
\citet{paulsen2012inexorable}, for example, used simulations to argue that in the early time of drop coalescence
a third regime beyond the Stokes and inertial regimes exists, where inertia, viscosity and surface tension forces are all important.
\citet{baroudi2015dynamics} employed the Lattice Boltzmann method to model the coalescence of two drops in both a saturated vapor phase and a non-condensable gas phase, showing
good agreement with experiments \citep{yao2005coalescence}, and 
\citet{xia2019universality} was able to resolve the transition region between the inertially-limited and viscously-limited drop-drop coalescence regimes.
More recently, \citet{baroudi2020effect} used the Cahn-Hilliard equation to show the effect of interfacial mass transport on inertial drop spreading on solids with varying wetting properties.
Most of these previous works, however, are restricted to systems with $Oh\ll1$ or $Oh\sim1$, as simulations for $Oh\gg1$ have proved too costly. In this work we address this gap by considering not only systems with $Oh\ll1$ and $Oh\sim1$, but also systems with $Oh \gg 1$.

\section{Methodology}
\subsection{Experimental setup}
\begin{figure}
    \centering
    \includegraphics[width=13cm]{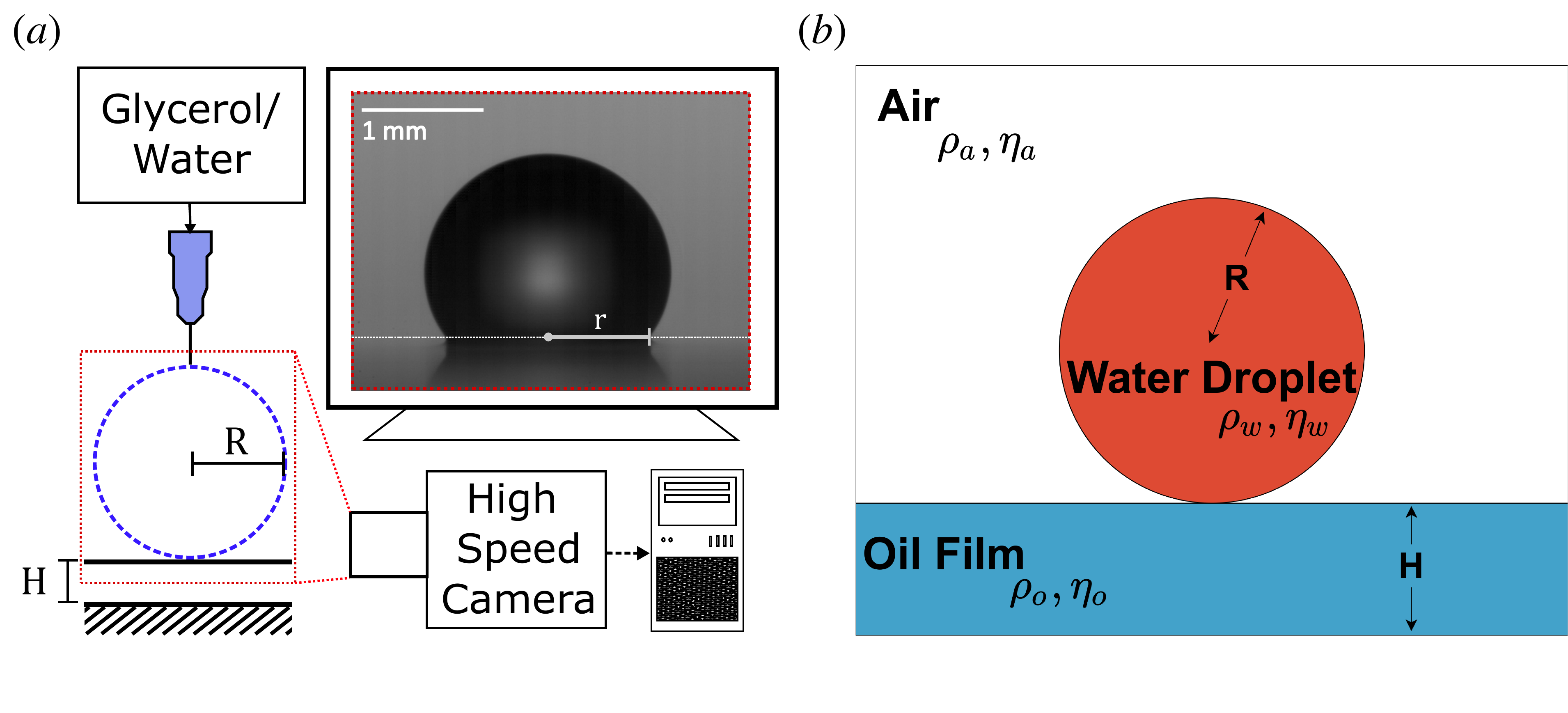}
     \caption{\label{fig1}(a) Experimental setup. Water/glycerol pendant drops of radius $R\approx 1$ mm and viscosity $\eta_w = [0.035 - 0.154]~ \textrm{Pa}\cdot \textrm{s}$ are brought into contact with a silicone oil film of height $H$ and viscosity $\eta_o = [0.33 - 1.54]~ \textrm{Pa}\cdot \textrm{s}$. Viscosity ratios $\eta_w:\eta_o$ are held at 1:10. The spreading radius $r$ is quantified using high-speed photography. (b) Initialization of the three phase flow simulations. The red phase represents the water drop with density $\rho_w$, viscosity $\eta_w$, and radius $R$. The blue phase represents the oil film with density $\rho_o$, viscosity $\eta_o$, and film thickness $H$. The colorless background represents air with density $\rho_a$ and viscosity $\eta_a$.}
\end{figure}

Figure \ref{fig1}(a) describes the experimental setup. Here a glycerol/water pedant drop of radius $R=1$ mm comes in contact with a quiescent silicone oil film of thickness $H$. Contact line spreading and engulfment dynamics were imaged using a Photron Fastcam SA5 with a Nikon 200mm f/4 AF-D Macro lens and two teleconverters (Nikon 2x and 1.7x) at 3000 fps. The spreading radius $r$ was tracked in Matlab using partial area subpixel edge detection techniques affording resolutions below $6 \mu$m/pixel \citep{trujillo2013accurate,trujillo2019code}. %

The initial drop radius is held constant and $Oh$ varied by changing the oil film viscosity $ \eta_o=[0.33- 7.6] \textrm{Pa}\cdot$s. This range of oil viscosities was created by mixing 20 cSt and 1000 St silicone oils. Drops were formed using mixtures of glycerol and DI water and controlled to maintain viscosity ratios $\eta_w/\eta_o\approx0.1$: $35$ mPa$\cdot$s, $68$ mPa$\cdot$s, $154$ mPa$\cdot$s and $705$ mPa$\cdot$s. All viscosities were characterized on an Anton-Paar 702 Rheometer with the parallel plate geometry from shear rates $0.1$ to $100$ 1/s, and all mixtures were confirmed to behave as Newtonian fluids. 

Film thickness ratios were varied from $H/R = [0.2- 4.4]$ with the drop radius held constant. 
For $H < 1$ mm, $H$ was controlled by depositing a prescribed mass of silicone oil measured via laboratory balance onto a glass slide cleaned with DI water and isopropyl alcohol. The slide was then placed on a level surface and the oil was allowed to fully spread until becoming pinned by the slide's corners and forming a level oil-air interface.
For $H \ge 1$ mm, glass slides were glued together with a hot glue gun to form a rectangular well. The well depth was controlled using machinery shims. A prescribed mass of oil was again added using a laboratory balance. To avoid complications in side view imaging from the concave meniscus formed by the oil on the glass wall, care was taken to ensure the desired film thickness was slightly larger than the well depth so imaging would occur on a convex meniscus not obstructed by the oil's contact line. 

The surface tension between the oil phase and the glycerol/water phase was measured to be approximately $\sigma_{ow} = 0.02 \textrm{N/m}$ by comparing the change in contact angle $\alpha$ of a sessile glycerol/water drop on a clean glass slide that is either surrounded by air or by silicone oil. The Young-Dupr\'e equation 
was then applied assuming the surface tension between the air and the glass is $0$ and that $\alpha = 0^\circ$ for oil on glass.

\subsection{Simulation setup and parameters}

Numerical simulations of the drop engulfment process, 
were performed in a rectangular domain, see Figure \ref{fig1}. A water/glycerol drop with radius $R$, density $\rho_w$, and viscosity $\eta_w$ is placed in contact with an oil film. The center of the drop is placed at $H+R+\delta$, where $H$ is the film thickness, $\delta$ the diffuse interface thickness, $\rho_o$ the density, and $\eta_o$ the oil's viscosity. Air, with density $\rho_ a$ and viscosity $\eta_ a$, makes up the background phase. Table \ref{Tab1} provides the relative magnitudes of the parameters and the range of parameters explored in the simulations and experiments. Densities and viscosities are fixed in the simulations: $\rho_w/\rho_a=\rho_o/\rho_a=100$; $\eta_o/\eta_a=10\eta_w/\eta_a=100$. The principal interest is to characterise the viscous and inertial engulfment dynamics for different ratios of $H/R$, which is achieved by exploring the phase space $Oh=[0.01-40]$ and $H/R=[0.03-4]$, see Table \ref{Tab1}.

We consider a ternary flow system that is composed of a water/glycerol drop, an oil film, and air, where the spreading coefficients determine the equilibrium contact angles \citep{guzowski2012structure,pannacci2008equilibrium}. The spreading factor can be calculated as follows: $S_o=\sigma_{wa}-\sigma_{wo}-\sigma_{ao}$. Since the surface tension of oil is typically smaller than water/glycerol, the silicone oil will perfectly wet the water/glycerol surface. Thus we only consider spreading coefficients $S_o>0$, where the oil fully engulfs the water/glycerol drop at equilibrium. 

Inherent to the phase field model are diffuse interfaces with finite thicknesses. Here we ensure that the interface thickness is small by using a Cahn number $Cn=\delta/R<0.05$, which achieves the sharp interface limit by performed convergence tests. Since we are interested drops with radii smaller than the capillary length, we are justified in neglecting the gravitational force in our simulations. From the simulations we extract the fluid velocity fields, the interfacial shapes, the position of the apparent contact line and the drop's center of mass calculated as $C_m=\sum\phi_w y/(\pi R^2 )$, where $y$ represents the vertical displacement scaled by the drop diameter $2R$.

A detailed comparison between 2D and 3D simulations is conducted in Appendix \ref{app1}, which shows that there are only minor differences between the two. Thus, we focus the article on 2D simulations using a $D2Q9$ lattice, as these give a good qualitative description of the flow.

 \begin{table}
\caption{\label{Tab1} Physical properties of the liquids used in ternary flow system.}
\centering
\begin{tabular}{ccc}
 Parameters &~~~~~ Description& Value \\ 
 \hline
    $\rho_o/\rho_a$  &~~~~~ density ratio      ~~~~~& 100    \\
    $\rho_o/\rho_w$  &~~~~~ density ratio    ~~~~~& 1    \\
    $\eta_o/\eta_a$  &~~~~~ viscosity ratio    ~~~~~& 1000    \\
    $\eta_o/\eta_w$  &~~~~~ viscosity ratio ~~~~~& 10    \\
    $Oh$             &~~~~~ Ohnesorge number                     ~~~~~& $0.01-40$\\
    $S_o$            &~~~~~ Spreading factor for oil              ~~~~~& >0            \\
    $H/R$            &~~~~~ film height to drop radius ratio         ~~~~~& $0.03-4$    \\
\end{tabular}

\end{table}

\subsection{Conservative phase field Lattice Boltzmann method}

The velocity-pressure LBM \citep{baroudi2021simulation} is utilized to solve the mass and momentum transport equations, and the conservative phase-field LBM is employed to capture the interfaces in order to simulate the three phase engulfment process in 2D and 3D. In the following equations, the order parameter $0\leq\phi_i\leq1$ is used to separate different phases, where $\phi_i=1$ represents the bulk region of component $i$ and $0<\phi_i<1$ represents the interface region. For order parameter $\phi_ i$, the discrete Boltzmann equation (DBE) can be expressed as
\begin{equation}\label{2:1}
\left(\frac{\partial}{\partial t}+\boldsymbol{e}_\alpha\cdot\nabla\right)h_\alpha^i=
   -\frac{1}{\lambda_\phi}(h_\alpha^i-h_\alpha^{i,eq})+{S_{\alpha}^i},
\end{equation}
where $h_\alpha^i$ represents the particle distribution function (PDF) for order parameter $\phi_i$ along the $\boldsymbol{e}_\alpha$ direction. $\boldsymbol{e}_\alpha$ denotes the discrete velocity vector in $D2Q9$ or $D3Q27$ lattice \citep{lee2005stable}, and $\lambda_\phi$ is the relaxation time. $\Gamma_\alpha$ takes the form:
$$\Gamma_\alpha=t_\alpha\left[1+\left(\frac{\boldsymbol{e}_\alpha\cdot\boldsymbol{u}}{c_s^2}+\frac{\left(\boldsymbol{e}_\alpha\cdot\boldsymbol{u}\right)^2}{2c_s^4}-\frac{\boldsymbol{u}\cdot\boldsymbol{u}}{2c_s^2}\right)\right],$$  
where $t_\alpha$ is the weight, $c_s$ is the speed of sound, and $\boldsymbol{u}$ is the macroscopic velocity vector of the flow field \citep{he1997theory}. The equilibrium distribution equation $h_\alpha^{i,eq}$ can be calculated as 
\begin{equation}
    h_\alpha^{i,eq}=\phi_i\Gamma_\alpha.
\end{equation} 
 The last term, $S_\alpha^i$, is the intermolecular forcing term
 \begin{equation}
{S}_\alpha^i=\Gamma_\alpha(\boldsymbol{e}_\alpha-\boldsymbol{u})\cdot\left(\frac{4 }{\delta}\frac{\nabla\phi_i}{|\nabla\phi_i|}\phi_i(1-\phi_i)-\frac{\phi_i^2}{\sum_{j=1}^3\phi_j^2}\sum_{j=1}^3\frac{4 }{\delta}\frac{\nabla\phi_j}{|\nabla\phi_j|}\phi_j(1-\phi_j)\right),
\end{equation} 
where $\delta$ denotes the finite interface thickness.

The DBE for mass and momentum transport can be expressed as
\begin{equation}\label{2:2}
    \left(\frac{\partial }{\partial t}+\boldsymbol{e}_\alpha \cdot\nabla\right)g_\alpha =-\frac{1}{\lambda}(g_\alpha-g_\alpha^{eq})+F_\alpha,
\end{equation}
where $g_\alpha$ represents the PDE for the pressure, with the function
\begin{equation}\label{Ceq:16}
g_\alpha^{eq}=t_\alpha\bar{p}+\Gamma_\alpha c_s^2 -t_\alpha c_s^2,
\end{equation}
where $\bar{p}=P/\rho$. $P$ represents the dynamic pressure and $\rho$ the mixture density. $F_\alpha$ represents the forcing term
\begin{multline}
    F_\alpha=-\Gamma_\alpha (\boldsymbol{e}_\alpha-\boldsymbol{u})\cdot\left(\frac{1}{\rho}\nabla P\right)+\Gamma_\alpha(0)(\boldsymbol{e}_\alpha-\boldsymbol{u})\cdot\left(\nabla \bar{p}\right)+\\\Gamma_\alpha (\boldsymbol{e}_\alpha-\boldsymbol{u})\cdot\left[\frac{\nu}{\rho}(\nabla\boldsymbol{u}+\nabla\boldsymbol{u}^T)\nabla\rho+\frac{1}{\rho}\boldsymbol{F}_s\right],
\end{multline}
where $\nu$ is the kinematic viscosity and $\boldsymbol{F}_s$ the surface tension force. In our simulations, the continuum surface tension force model is employed \citep{kim2005continuous}:
$$\boldsymbol{F}_{s}=-\frac{3\delta}{2}\sum_i\sigma_i\nabla\cdot\left(\frac{\nabla\phi_i}{|\nabla\phi_i|}\right)   |\nabla\phi_i|  \nabla\phi_i,$$  
where $\sigma_i=\left(\sigma_{ij}+\sigma_{ik}-\sigma_{jk}\right)/2$. The gradient terms of the macroscopic values which appear above can be calculated by the second order isotropic finite difference method \citep{lee2005stable}.

The governing equations recovered through the Chapman-Enskog expansion are
\begin{equation}
\frac{\partial \bar{p}}{\partial t}+\boldsymbol{u}\cdot\nabla \bar{p}+ c_s^2 \nabla\cdot \boldsymbol{u}=0,
\end{equation}
\begin{equation}
\frac{\partial \boldsymbol{u}}{\partial t}+\nabla\cdot(\boldsymbol{uu})=-\frac{1}{\rho}\nabla P+\frac{1}{\rho}\nabla\cdot \eta\left(\nabla\boldsymbol{u}+(\nabla\boldsymbol{u})^T\right)-\frac{3\delta}{2\rho}\sum_{i=1}^{3}\sigma_i\nabla\cdot\left(\frac{\nabla\phi_i}{|\nabla\phi_i|}\right)   |\nabla\phi_i| \nabla\phi_i,
\end{equation}
\begin{equation}
 \frac{\partial\phi_i}{\partial t}+\nabla\cdot(\phi_i\boldsymbol{u})= \nabla\cdot M\left(\nabla\phi_i-\frac{4}{\delta}\frac{\nabla\phi_i}{|\nabla\phi_i|}\phi_i(1-\phi_i)
    +\frac{\phi_i^2}{\sum_{j=1}^3\phi_j^2}\sum_{j=1}^3\frac{4}{\delta}\frac{\nabla\phi_j}{|\nabla\phi_j|}\phi_j(1-\phi_j)\right).
\label{Ceq:3}
\end{equation}
and the density and viscosity defined for the three phases
\begin{equation}
    \rho=\sum_{i=1}^3\rho_i\phi_i
\end{equation}
\begin{equation}
    \eta=\sum_{i=1}^3\eta_i\phi_i
\end{equation}
At the boundaries of the domain, we use symmetric boundary conditions to the left and right boundaries, while the no-slip boundary condition is applied to the top and bottom boundaries. The details of the LBM derivation and Chapman-Enskog expansion can be found in appendix \ref{C-E}.

\section{Results and Discussion}

\subsection{Inertial-capillary drop engulfment $Oh<1$}

\begin{figure*}
\centering
  \includegraphics[width=\linewidth]{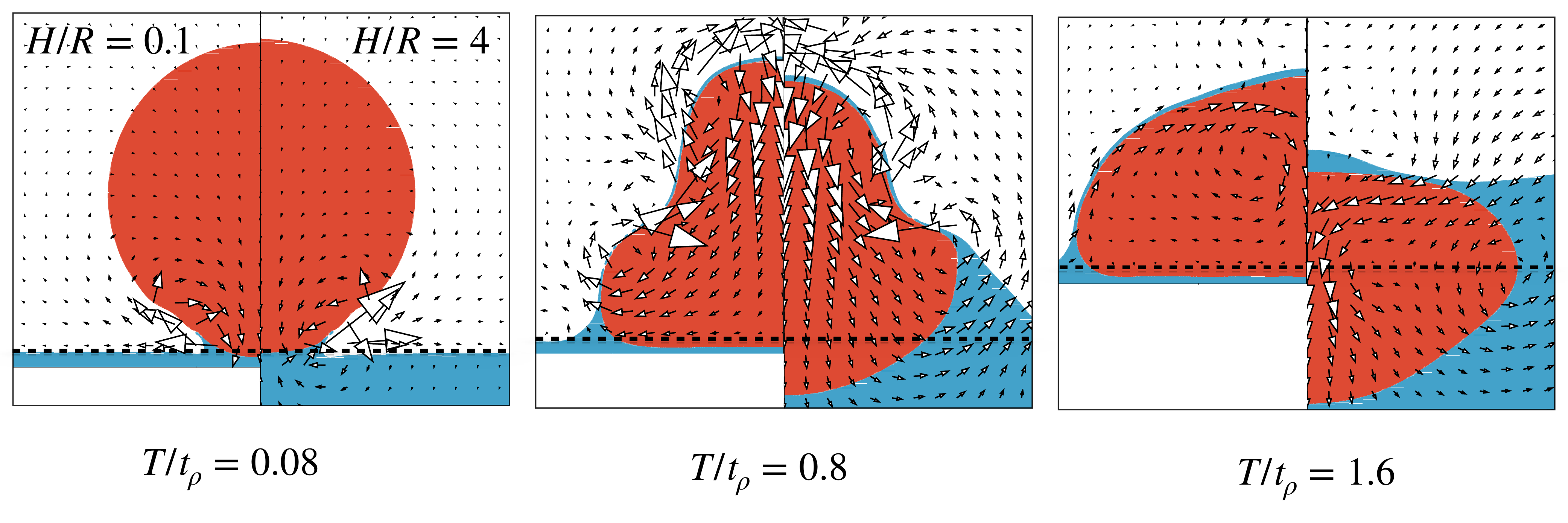}
    \caption{\label{fig2} Three snapshots in time of simulations of a glycerol/water drop engulfed by a thin $H/R=0.1$ (left panel) and a thick $H/R=4$ (right panel) oil film in the inertial regime, $Oh=0.07$. The contours indicate the three different phases, air (colorless), droplet phase (red) and oil film (blue). Black arrows show the flow field's velocity vectors. Black dash lines indicate the initial oil film height. Post drop-film contact the drop deforms and oscillates, where the droplet is quickly engulfed by the oil. As the film thickness increases capillary waves are formed on the film's interface and affect the drop dynamics.
    }
\end{figure*}

Figure \ref{fig2} shows the 2D simulation results of a water drop contacting an oil film in the inertial-capillary regime, $Oh \ll 1$,
 for varying film thicknesses $H/R$. Immediately post contact, the interface near the the oil-water-air contact line deforms creating a large curvature tangential to the wall that forms an apparent spreading radius. This curvature drives the apparent spreading of the drop. Post spreading the drop becomes engulfed by the oil film and either sinks into the film adopting a circular interface or is supported from below by the wall with an interface reminiscent of a spherical-cap.

Evident from Figure \ref{fig2} with $Oh=0.07$ are the striking morphological variations post drop-film contact in the inertial-capillary regime.
Here, the contours of the three different liquid phases and the velocity vectors are shown for $H/R=0.1$ (left panel) and $H/R=4$ (right panel).  
We notice that in the short time, $T/t_{\rho}=0.08$, the drop's interfacial shape for the two film heights appears identical with a large velocity generated near the three phase line as the oil spreads onto the drop.
A region of large curvature is generated by the oil film as it spreads on the drop and the subsequent capillary force pulls the drop in the direction tangential to the wall with an apparent spreading radius. 
This rapid spreading of the drop causes the interface to significantly deform and oscillate, a result of the lack of viscous resistance to motion in the inertial-capillary regime. In a relatively short time the oil completely engulfs the drop, inducing a capillary force that
pushes the drop towards the substrate. A squeeze flow in the gap between the drop interface and the wall is generated, with a build up of pressure that is naturally a function of the gap height (Figure \ref{fig2} $T/t_{\rho}=0.8$). 
The difference in the flow induced pressure underneath the drop significantly affects the drop's interfacial dynamics for $T/t_{\rho}=0.8-1.6$, where 
for thin films $H/R=0.1$ the drop extends along the wall and for thick films $H/R=4$ the drop is transported into the film with capillary waves formed on the oil interface.
At long times, not shown here but implied by Figure \ref{fig2}, the drop adopts a spherical cap shape for the thin film case with a meniscus at the apparent spreading radius filled with the oil phase. 
In contrast, for thick films the drop is transported into the film and relaxes into a circular drop below a flat oil interface. 
As there are no buoyancy effects in the simulations, only viscous resistance will halt the drop's motion, which in the inertial regime is small. 

\subsubsection{Droplet center of mass motion and apparent spreading dynamics}
\begin{figure}
  \includegraphics[width=0.6\linewidth]{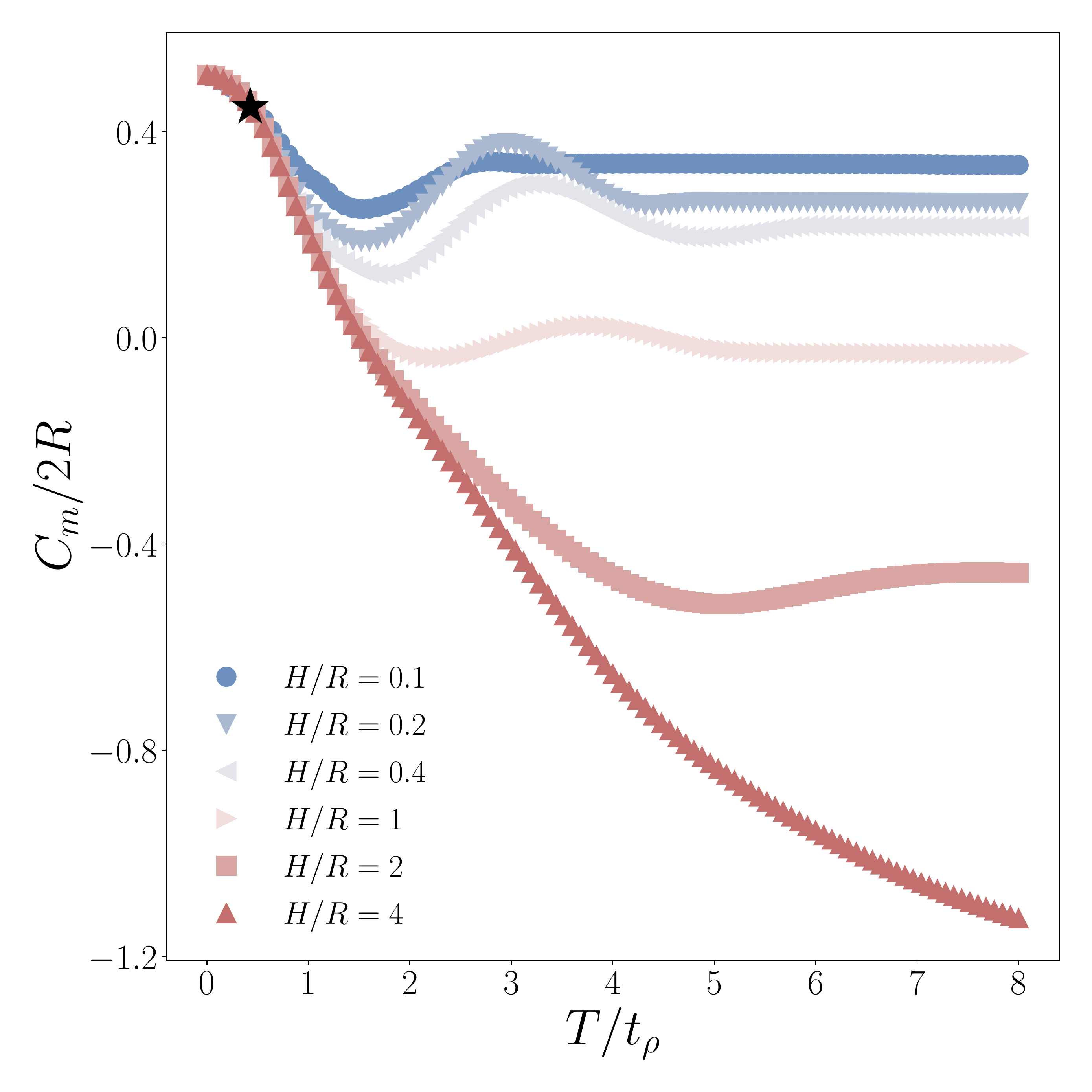}
  \centering
\caption{The scaled center of mass ($C_m/2R$) position (wall normal) of the drop as a function of time post contact with an oil film of varying thickness ($H/R=[0.1- 4]$
, ($\mycircle{black}$,$\blacktriangledown$,$\blacktriangleleft$,$\blacktriangleright$,$\mysquare{black}$, $\blacktriangle$)) 
for $Oh=0.07$. The drop is fully engulfed by the oil film at $T/t_\rho\approx0.48$, and we mark the engulfment time in the figure ($\bigstar$). Post engulfment the large inertia causes an oscillatory motion of the drop's center of mass for thin films with oscillations damped when $H/R\approx 4$.
}\label{coms}
\end{figure}

To characterise the drop's engulfment dynamics we extract the position of the center of mass $C_m$ in the wall normal direction relative to the position of the initial film-air interface
and the initial drop radius $R$ from the simulations. Here the center of mass $C_m$ is used as a measure for the drop's motion into the oil film. In Figure \ref{coms}, we compare the evolution of the drop's center of mass $C_m$ for oil films of different thicknesses. 
Initially,
there appears to exist a universal short-time regime that is independent of the film height for $T/t_\rho<0.2$.
This collapse to the same curve indicates that in the short-time, the spreading of the drop drives the displacement of the center of mass. At long times, the position of the center of mass $C_m$ is highly affected by the ratio $H/R$, with larger film thickness $H/R$ allowing for larger wall normal displacements. 


A wave-like motion of $C_m$ appears for $H/R\leq 1$ in Figure \ref{coms}, which is formed by an inertial-capillary wave caused by the imbalance between the drop's initial shape and its equilibrium state \citep{ding2012propagation}. The inertial-capillary wave motion can also be seen in Figure \ref{fig2} when $H/R=0.1$ as the drop oscillates up and down between time $T/t_{\rho}=0.8-1.6$. This wave-like motion becomes less pronounced as the film layer thickness increases and when the oil coating is sufficiently thick $H/R>4$, no inertial-capillary wave is formed. The pressure build up by the squeeze flow between the drop interface and the wall ultimately arrests the drops at different positions in the film as a function of $H/R$.


\begin{figure}
  \centering
  \includegraphics[width=\linewidth]{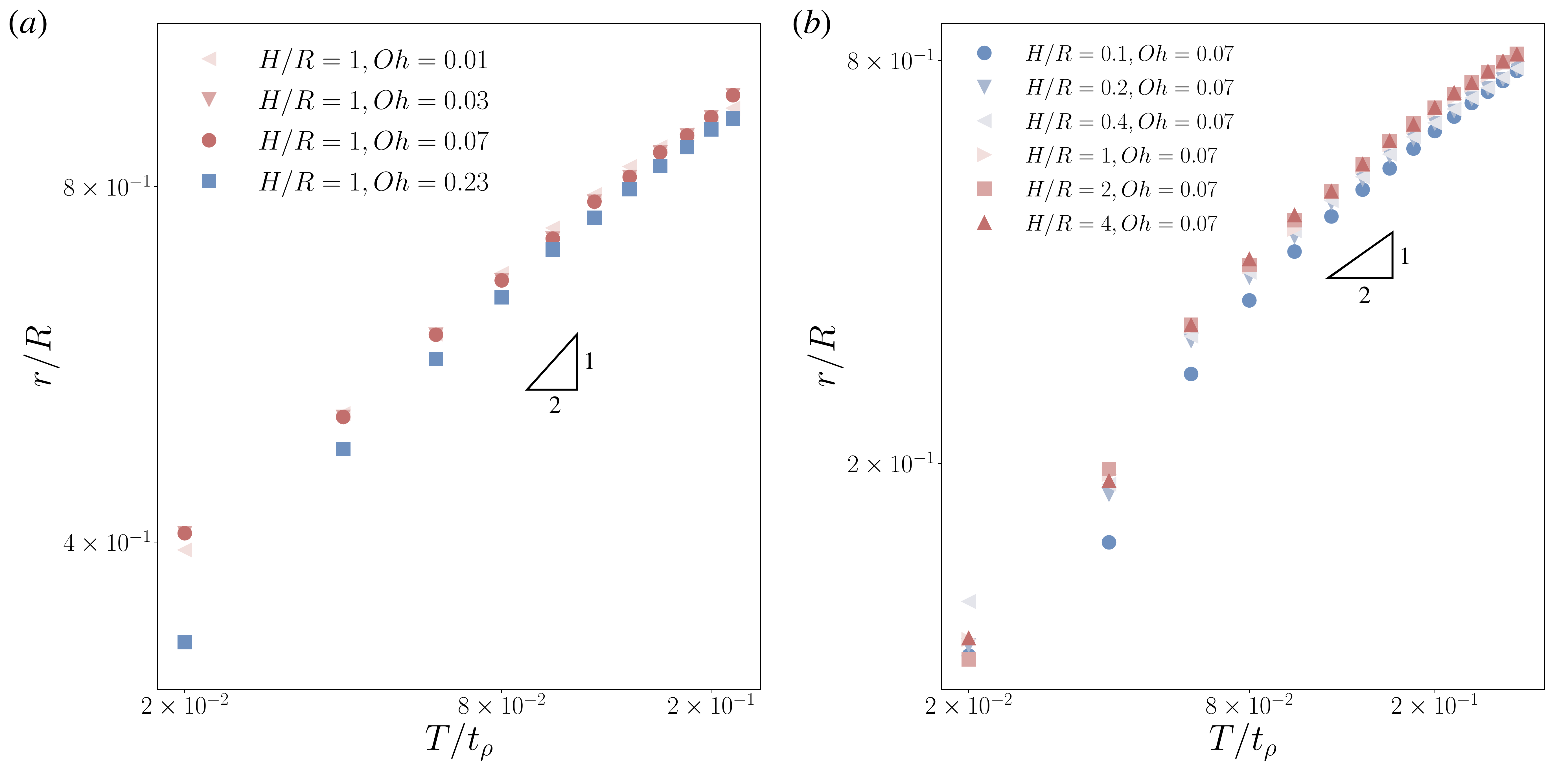}
\caption{\label{fig6} {The apparent contact-line radius $r/R$ plotted against time $T/t_{\rho}=[0-0.2]$ for (a) $Oh=[0.01-0.23]$
($\blacktriangleleft$,$\blacktriangledown$,$\mycircle{black}$, $\mysquare{black}$) 
and $H/R=1$ and (b) $Oh=0.07$ and $H/R=[0.1- 4]$
($\mycircle{black}$,$\blacktriangledown$,$\blacktriangleleft$,$\blacktriangleright$, $\mysquare{black}$,$\blacktriangle$). 
Reference triangles indicate a slope $\beta=0.5$. 
Regardless of $Oh$ and $H/R$ it can be seen that $r/R\sim(T/t_\rho)^{1/2}$}, analogous to the coalescence of two drops in the inertial regime.
}
\end{figure}

To describe the dynamics we also extract the apparent spreading radius $r$ from the simulations. In the inertial-capillary regime, the log-log representation of the apparent spreading radius for $Oh=[0.01-0.23]$ and $H/R = 1$ is shown in Figure \ref{fig6}(a) and for $Oh=0.07$ and $H/R= [0.1-4]$ in Figure \ref{fig6}(b). 
Here we find that in this apparent spreading regime that the apparent spreading radius is insensitive to the film height, as was the drop's center of mass $C_m$. Further, the simulations of Figure \ref{fig6} indicate that $r$ has a power-law dependence with time as $r/R\sim({T/t_\rho})^{1/2}$, which is consistent with miscible inertial drop-film coalescence and inertially-limited engulfment of a water drop on a very thin oil film \citep{eggers1999coalescence,carlson2013short}. The differences between simulations on various film thickness $H/R=[0.1-4]$ appear to only be present at long times and are associated with the drop being transported by capillarity into the film.

\subsection{Visco-capillary drop engulfment $Oh>1$}\label{sec:ViscSpreading}

\begin{figure*}
\centering
  \includegraphics[width=\linewidth]{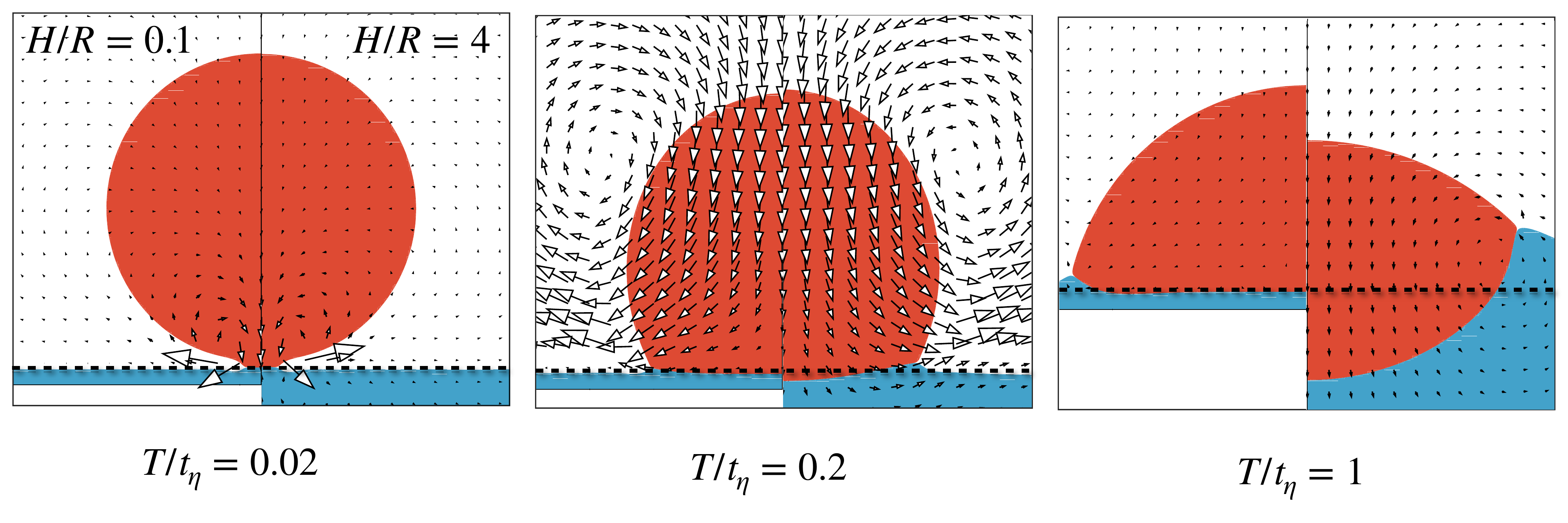}
    \caption{\label{fig2_visco} Simulations of glycerol/water drop spreading and being engulfed by a thin $H/R=0.1$ (left panel) and thick $H/R=4$ (right panel) oil film in the viscous regime, $Oh=5.27$. Black arrows illustrate the flow field's velocity vectors. Black dash lines indicate the initial oil film height. Post contact the drop spreads on the film prior to engulfment. It can be seen that regardless of film thickness, the drop's contact-line spreads at the same rate with similar drop morphologies at each time step. 
    }
\end{figure*}

Next we move on to study the less understood visco-capillary regime $Oh>1$. Figure \ref{fig2_visco} shows simulations of viscously-limited drop engulfment dynamics when $Oh=5.27$, for two different film thicknesses $H/R=0.1$ and $H/R=4$. 
Post drop-film contact, the velocity vectors mostly appear near the contact region, indicating that at short times the region of high curvature formed upon contact drives the radial spreading of the drop. 
Compared to the $Oh=0.07$ case, the drop's apparent spreading is slower, and the engulfment process is free of any inertial-capillary waves. 
Despite the fact that the capillary driving force is the same, any waves are effectively damped by the viscosity \citep{cuttle2021engulfment}. 
Comparing the simulation for different film thickness at long times shows that the drop can be fully engulfed for thick films, whereas when the oil film is very thin, the velocity inside the film becomes small and the simulations are unable to reach the fully engulfed state expected at equilibrium.

\subsubsection{Droplet center of mass motion and apparent spreading dynamics}
\begin{figure}
  \includegraphics[width=0.6\linewidth]{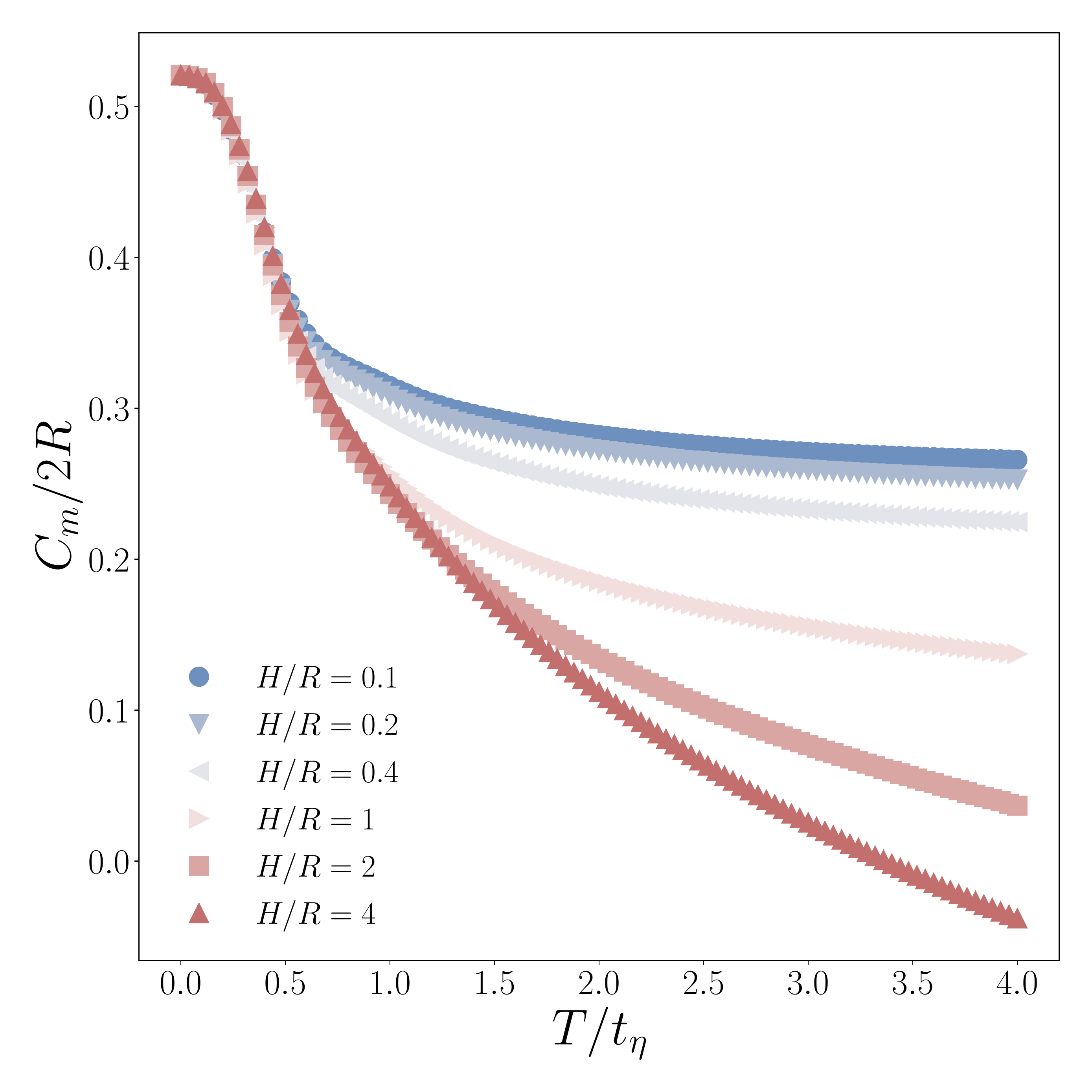}
  \centering
\caption{The scaled center of mass ($C_m/2R$) position (wall normal) of the drop as a function of time post contact with an oil film of varying thickness ($H/R=[0.1- 4]$
, ($\mycircle{black}$,$\blacktriangledown$,$\blacktriangleleft$,$\blacktriangleright$,$\mysquare{black}$, $\blacktriangle$)) 
for $Oh=5.27$. 
Drop spreading ends at $T/t_\eta=0.6$. As $H/R$ increases, the drop is able to sink further into the film.
}\label{coms_visco}
\end{figure}

In Figure \ref{coms_visco}, we compare the evolution of the drop's center of mass $C_m$ in the visco-capillary regime ($Oh>1$) for oil films of different thicknesses. 
As in the inertia-capillary regime,
the visco-capillary regime displays a universal short-time dynamics that are independent of the film height for $T/t_\rho<0.7$ indicating that the spreading of the drop is driving the displacement of the center of mass. At long times, the position of the center of mass $C_m$ is highly affected by the ratio $H/R$, with larger film thicknesses $H/R$ allowing for larger wall normal displacements. Variations in the thickness of the film show a diminishing effect on the engulfment process other than for the final position of the center of mass $C_m$. Here, no oscillations of the drop's center of mass are present. After complete engulfment the drop is arrested by the viscous force and reaches equilibrium.



\begin{figure}
  \centering
  \includegraphics[width=\linewidth]{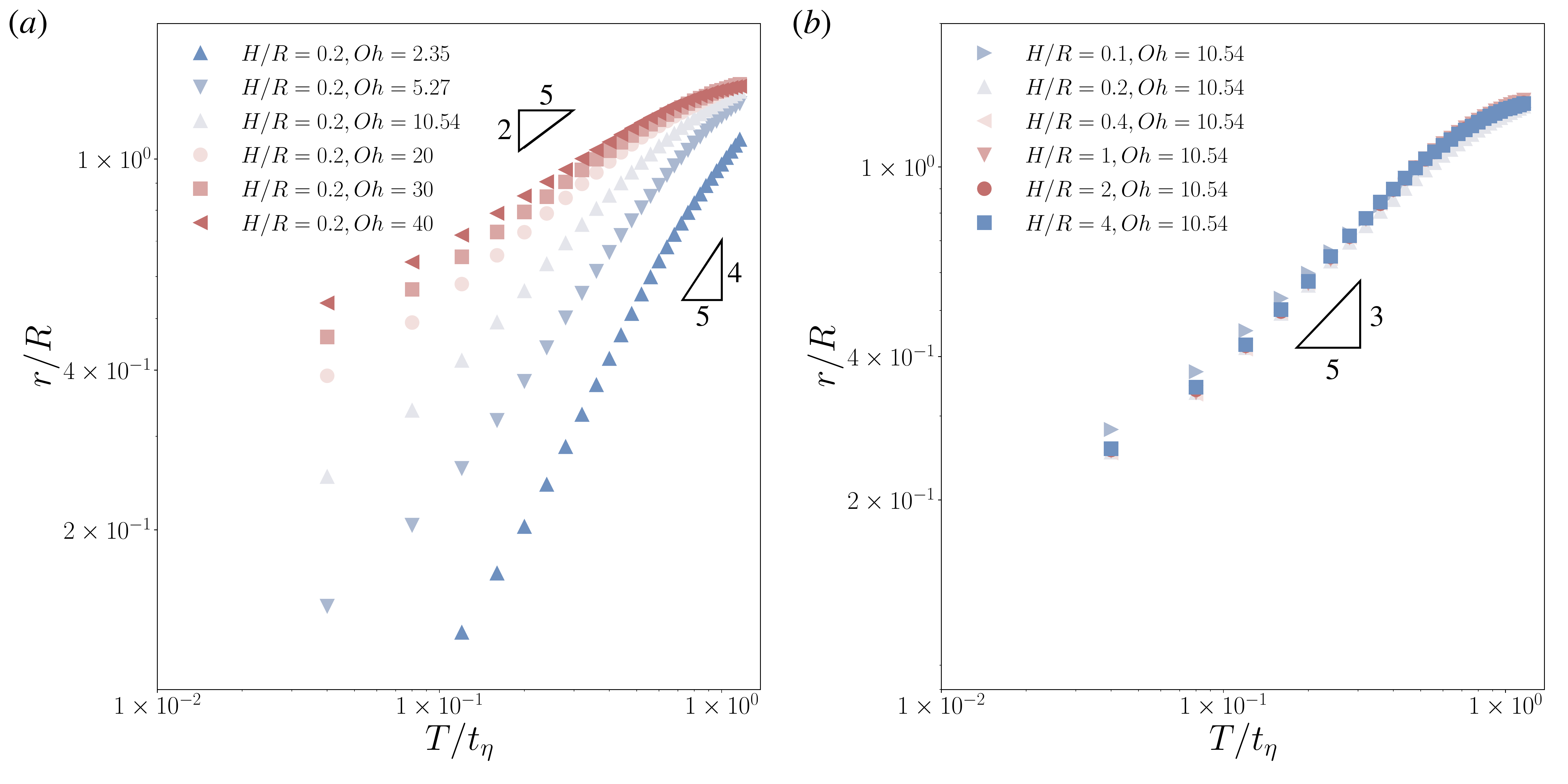}
\caption{The radius of the apparent contact-line $r/R$ against time
$T/t_\eta=[0-1]$ for (a) $Oh=[ 2.35 ,  5.27,  10.54 , 20, 30, 40]$ ($\blacktriangle$,$\blacktriangledown$,$\blacktriangle$,$\mycircle{black}$, $\mysquare{black}$,$\blacktriangleleft$) with film thickness $H/R=0.2$ and (b) $Oh=10.54$ with various film thickness $H/R=[0.1- 4]$  ($\blacktriangleright$,$\blacktriangle$,$\blacktriangleleft$,$\blacktriangledown$,$\mycircle{black}$,$\mysquare{black}$). Reference triangles highlight the power-law $\beta$ dependence with time. In (a), $\beta$ converges to $2/5$ from $4/5$ as $Oh$ increases from $Oh=2.35$ to $Oh=40$. In (b), $\beta$ is invariant to the film thickness, $H/R = [0.1-4]$, for constant $Oh$. }\label{vis}
\end{figure}

Figure \ref{vis} shows the log-log representation of the apparent spreading radius $r$ with $Oh=[1-40]$. 
In Figure \ref{vis}(a) we see that the apparent radius $r$ has a power-law dependence with time with an exponent that decreases with increasing $Oh$. 
This variation in the power-law exponent is a consequence of being in a transitional regime $Oh\sim1$, where there can still be some inertial effects \citep{paulsen2012inexorable}. However, as $Oh$ continues to increase, the exponent of the power-law appears to approach a constant value as $r/R\sim (T/t_\eta)^{2/5}$, which is comparable to the previous observations for water drop engulfment by a thin viscous oil film \citep{carlson2013short}. As in the inertial regime, we can see that the short-time dynamics are insensitive to the pre-wetted film height, as shown in Figure \ref{vis}(b), with a power law dependence of $r/R\sim (T/t_\eta)^{3/5}$ for $Oh = 10.54$.


\begin{figure*}
  \centering
  \includegraphics[width=\linewidth]{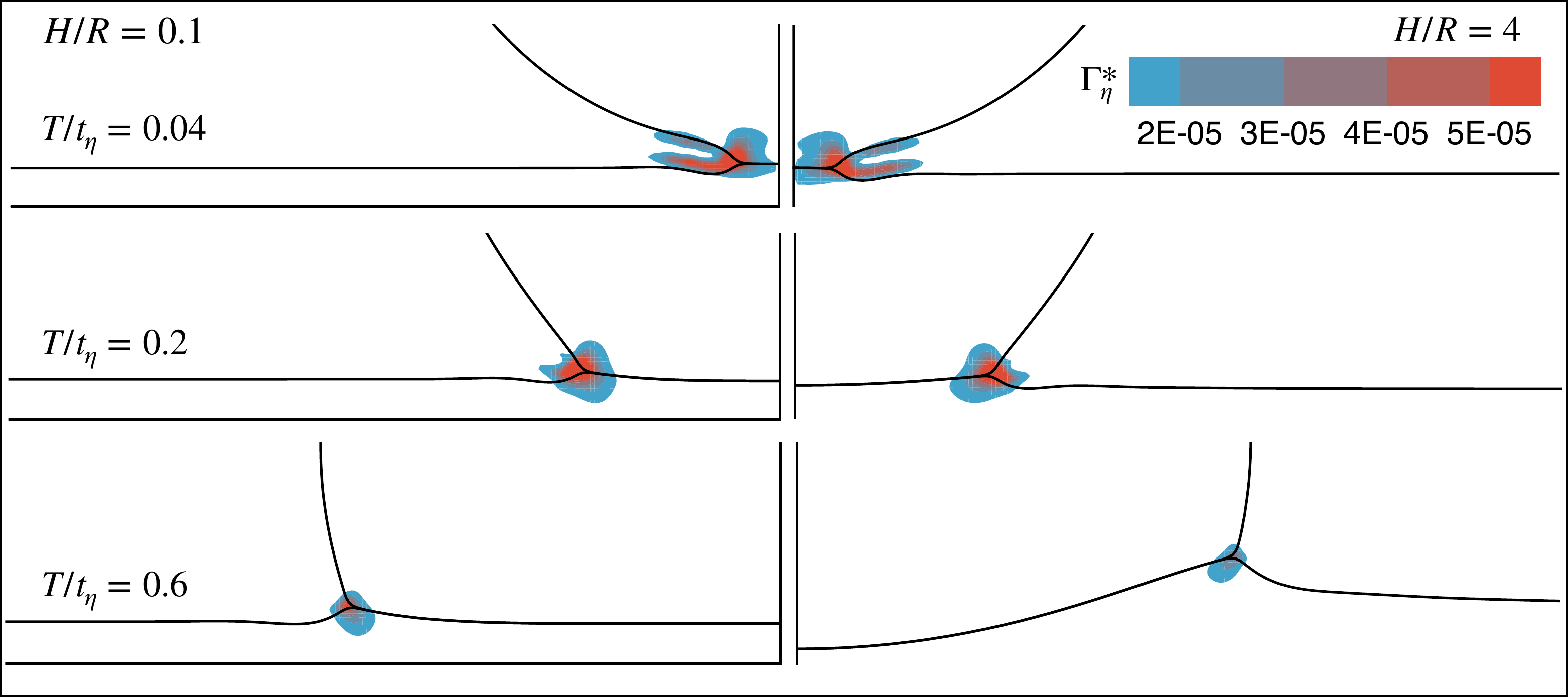}
    \caption{\label{ap3} Contour plot of the viscous dissipation $\Gamma_\eta^*$ of a thin film $H/R=0.1$ (left panel), and a thick film $H/R=4$ (right panel) for $T/t_\eta=0.04$; $T/t_\eta=0.2$; $T/t_\eta=0.6$, and $Oh=5.27$. The viscous dissipation is localized at the contact line region for thin and thick oil films, and gradually decreases.}
\end{figure*}

\begin{figure*}
  \centering
  \includegraphics[width=\linewidth]{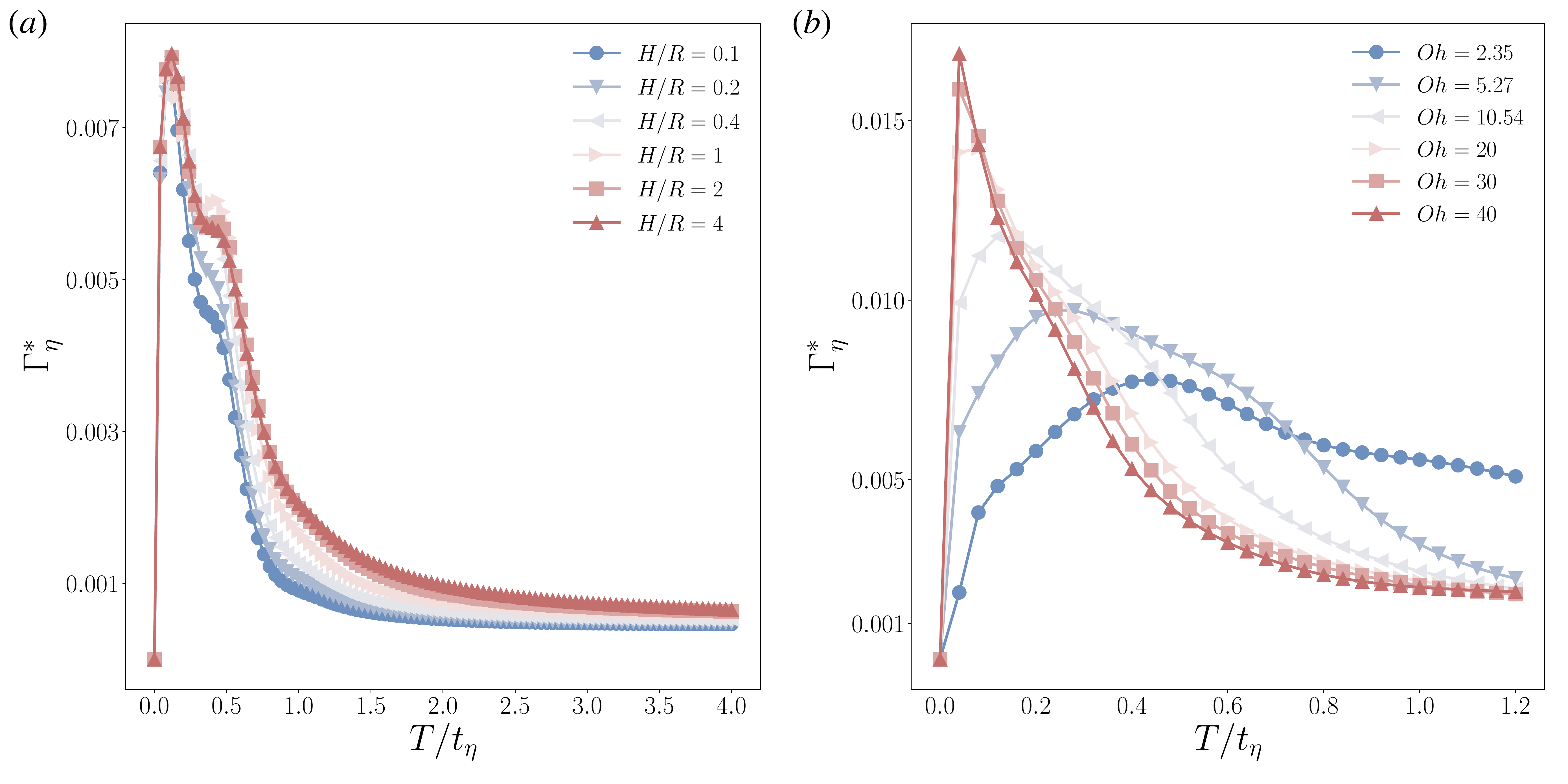}
\caption{\label{vcom} (a) The rate of the viscous dissipation as a function of time for $Oh=5.27$ with $H/R=[0.1-4]$, ($\mycircle{black}$,$\blacktriangledown$,$\blacktriangleleft$,$\blacktriangleright$,$\mysquare{black}$, $\blacktriangle$). (b) The rate of viscous dissipation as a function of time for $Oh=[2.35-40]$, ($\mycircle{black}$,$\blacktriangledown$,$\blacktriangleleft$,$\blacktriangleright$,$\mysquare{black}$, $\blacktriangle$). (a) For varying film thickness, the viscous dissipation first rapidly increases while the drop spreads on the oil film due to the large initial curvature near the three phase line and gradually dissipates over time as viscosity diffuses the energy. The film thickness shows little effect on the rate of dissipation for constant $Oh$.
(b) The viscous dissipation is depressed for $Oh\approx 1$ and converges to a similar trend as $Oh$ approaches $40$. This variation in dissipation over time as $Oh$ increases accounts for the changing power-law evolution of the apparent contact line during spreading from $\beta=4/5$ to $\beta=2/5$. 
}

\end{figure*}

To better understand the invariance of the power-law dependence of the spreading radius in time to the film thickness, we would like to compare the dissipation over time when $Oh=5.27$ for different $H/R$. Here the viscous dissipation $\Gamma_\eta$ of the whole simulation domain can be formulated as
\begin{equation}
    \Gamma_\eta=\int_\Omega \eta(\phi)\nabla u:(\nabla u +\nabla^T u)d\Omega.
\end{equation}
It can then be scaled by the surface energy $E_s=\sigma_{ow} U R$, where the capillary speed $U=\sigma_{ow}/\eta_w$ is considered as the reference velocity \citep{carlson2011dissipation}. The dimensionless form of the viscous dissipation is denoted as $\Gamma_\eta^*=\Gamma_\eta/E_s$. 
 
Figure \ref{ap3} shows the viscous dissipation per volume for thin $H/R = 0.1$ and thick $H/R = 4$ films at times ranging from $T/t_{\eta} = [0.04 - 0.6]$ for $Oh = 5.27$. It is clear that over the course of the initial spreading regime, the majority of the dissipation is centered around a small area in the region of high curvature near the triple contact point, regardless of the film thickness. This invariance to film thickness in the profile of the viscous dissipation supports the prior observation of the invariance in the power-law exponent for varying film thicknesses, Figure \ref{vis}(b).

Figure \ref{vcom}(a) shows the rate of the total viscous dissipation for varying film thickness $H/R = [0.1 - 4]$ with $Oh = 5.27$. 
Here we see that the rate of viscous dissipation throughout spreading $T/t_{\eta}<1$ is approximately constant, with only slight deviations during engulfment $T/t_{\eta}>1$ owing to the variations in the film thickness, supporting again the invariance of the power-law exponent of Figure \ref{vis}(b) to film thickness.
Additionally, the first peak ends at $T/t_\eta\approx1$, which is consistent with the transition from apparent spreading to engulfment.
It can be seen that a thinner film will overall have less viscous dissipation, indicating that the flow structure at the contact line is smaller than the pre-wetted film thickness. Figure \ref{vcom}(b) shows the rate of the scaled viscous dissipation for various $Oh$ with $H/R = 0.2$. When $Oh$ is large, i.e. $Oh>10$, we see that the rate of viscous dissipation over time converges to a single trend and becomes insensitive to $Oh$, paralleling the convergence of the power-law exponent to $2/5$ in Figure \ref{vis}(a).
 When the Ohnesorge number is in the transition regime, i.e. $Oh=[1-10]$, the rate of dissipation during spreading varies with $Oh$, resulting in the varying power-law exponents of Figure \ref{vis}(a).


\begin{figure}
	\centering
  \includegraphics[width=0.5\linewidth]{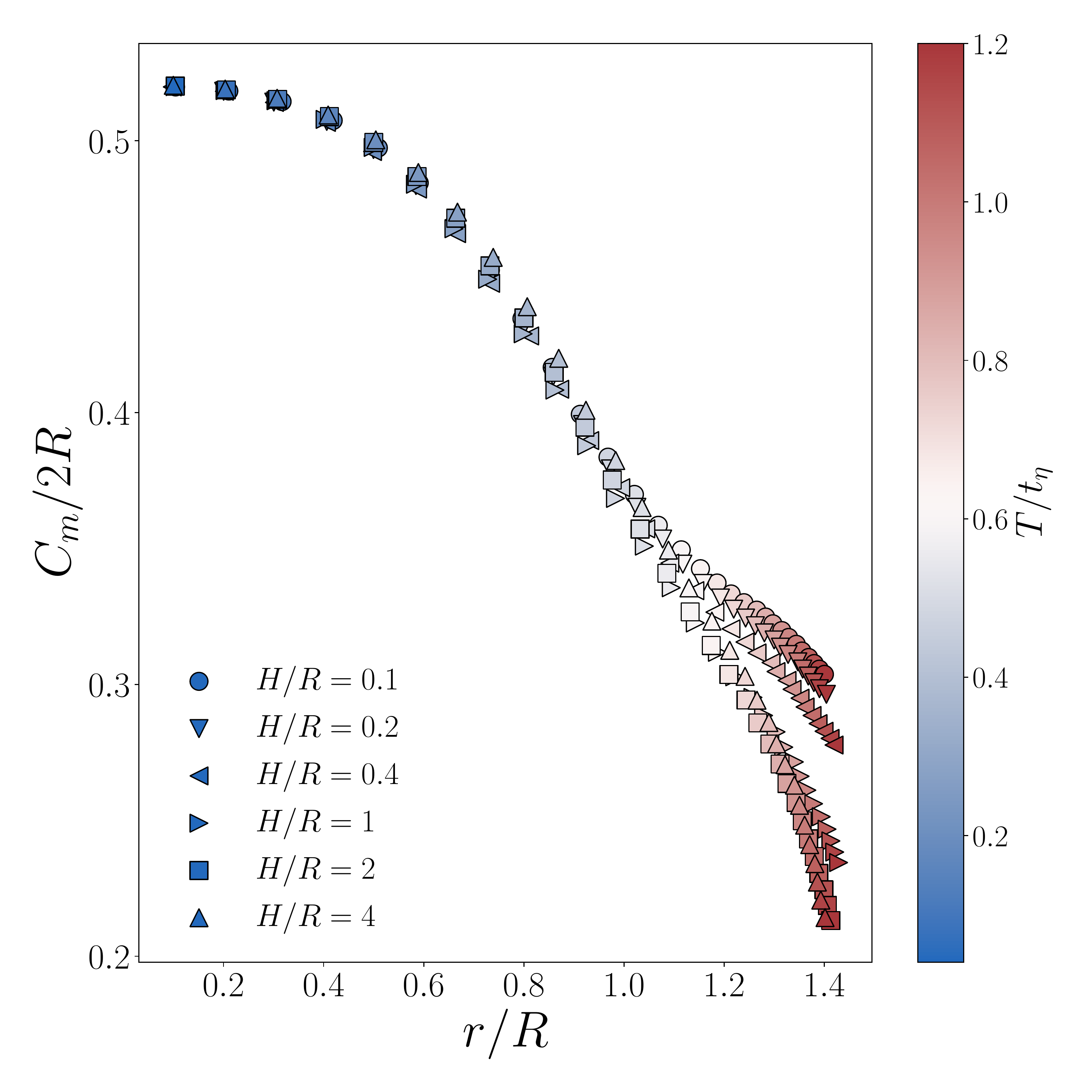}
    \caption{\label{mvr} The center of mass (wall normal direction) plotted against the spreading radius for $Oh=5.27$, $H/R=[0.1-4]$,($\mycircle{black}$,$\blacktriangledown$,$\blacktriangleleft$,$\blacktriangleright$,$\mysquare{black}$, $\blacktriangle$). During short-time spreading, $T/t_\eta<0.6$, the position of the drop's center of mass with respect to its contact-line radius is quite similar for varying film thickness constant $Oh$. After $T/t_\eta=0.6$, the drop is fully engulfed by the oil film begins sinking into the film. Here the effect of the film thickness becomes apparent as it affects the pressure difference between the air and the oil film.
    }
\end{figure}

Figure \ref{mvr} shows the variation in the center of mass of the drop against the apparent spreading radius in the visco-capillary regime with $Oh = 5.27$.
It is obvious that the short-time spreading radius evolution defined as the rapid first stage engulfment in \citet{cuttle2021engulfment} are quite similar for different film thickness.
At later times, however, when $T/t_\eta>0.6$ and $r/R>1$, the differences in the engulfment dynamics for varying film thicknesses become obvious, with the center of mass being pulled into the film further and faster as the film thickness increases.
The large curvature of the film interface coupled with the large viscosity and localized viscous dissipation explain the invariance to film thickness during short-time spreading, $T/t_\eta<0.6$. 
This invariance to film thickness during rapid spreading further explains the invariance in the mass center evolution to film thickness, as it is the spreading that initially pulls the drop's center of mass towards the film.
As time progresses, however, the pressure difference becomes the only effect driving engulfment, allowing the center of mass to be pulled further into the film as the film thickness increases.


\subsection{Experimental considerations}

\begin{figure}
    \centering
    \includegraphics[width=\linewidth]{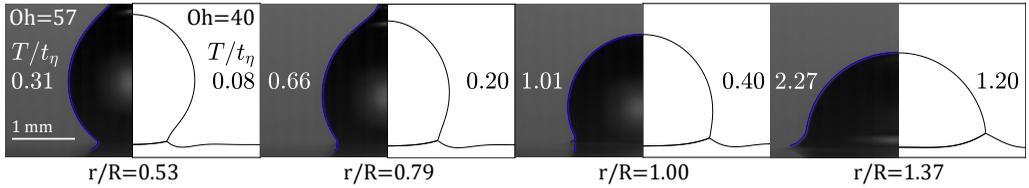}
     \caption{\label{fig:ohcompare}Comparison between the simulation and the experiment for $H/R=0.2$. Simulation: $Oh =40$. Experiment: $Oh = 57$, $\eta_o = 7.6$ Pa$\cdot$s, $\eta_w = 0.705$ Pa$\cdot$s. Snapshots are taken at similar spreading radii $r/R$. The simulation captures well the morphology of the drop's contour lines during the apparent spreading.}
\end{figure}

In order to validate the simulation results for visco-capillary drop engulfment from the previous section \S \ref{sec:ViscSpreading}, we focus our experiments in the regime when $Oh>1$. 
Figure \ref{fig:ohcompare} shows the results of a glycerol/water drop of radius $R \approx 1$ mm contacting a viscous oil film with a thickness of $H/R = 0.2$.
It can be seen qualitatively from Figure \ref{fig:ohcompare} with $Oh > 10$ that the 2D simulations capture well the shape of the liquid/gas interface during the initial spreading. 
However, discrepancies between the simulations and the experiments may arise from either the experimental drop's attachment to the needle tip or the simulation's lower dimensionality. The first two panels of Figure \ref{fig:ohcompare} show an obvious connection between the drop and the needle, which does influence the interfacial shape of the upper region of the drop. This influence, however, affects primarily the upper portion of the drop, leaving the region of high curvature near the 3-phase line unaffected. Additionally, in Appendix \ref{app1}, we perform the associated 3D simulations to test the dimensional dependency of the simulation during spreading and show that the apparent spreading dynamics can be reduced to a 2D problem. With these considerations in mind, we can meaningfully compare the results of the experiments with the 2D simulations previously shown.

\begin{figure}
    \centering
    \includegraphics[width = \linewidth]{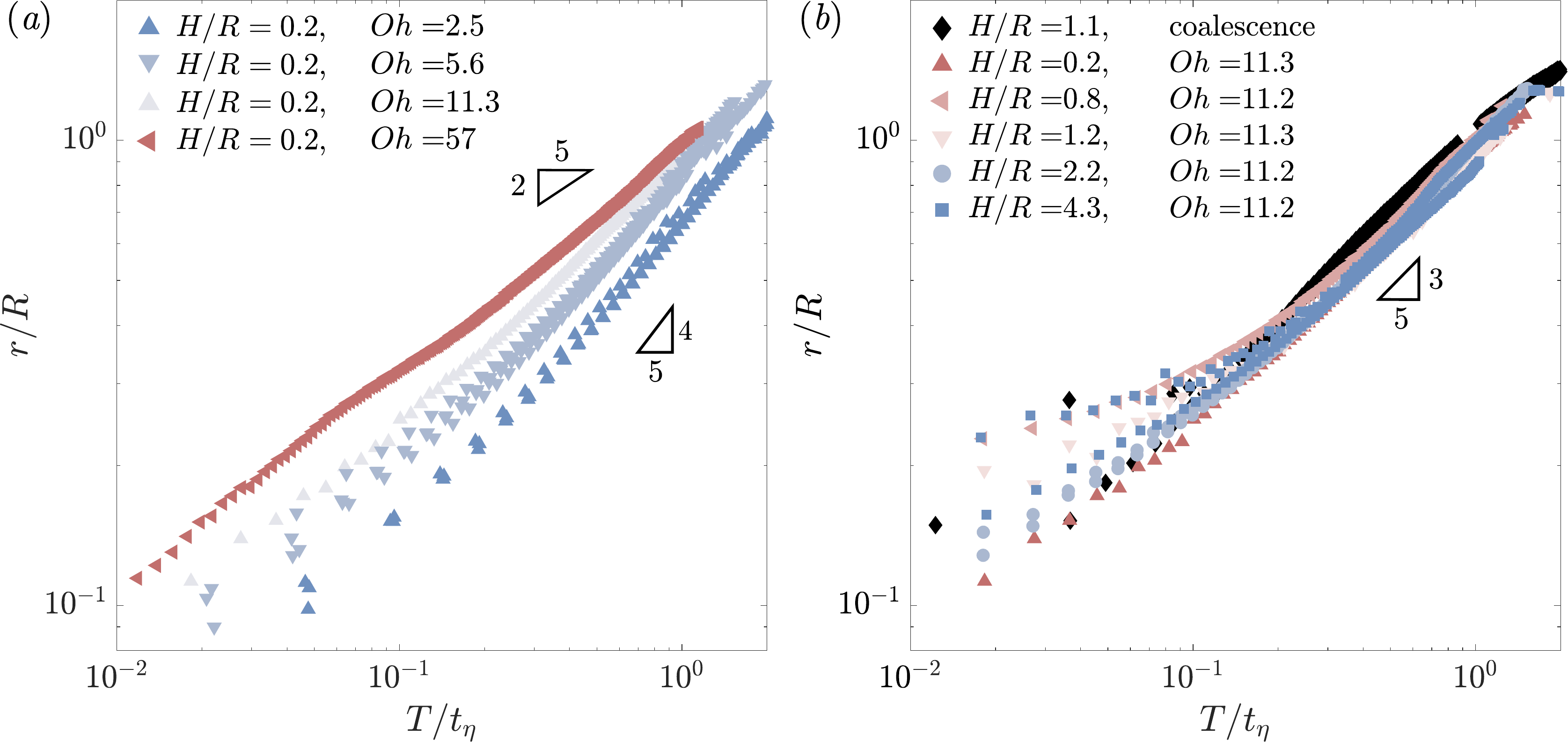}
    \caption{
    Experimentally measured apparent spreading radii $r/R$ against time $T/t_\eta$ for immiscible glycerol-water drops ($\blacktriangle$,$\blacktriangleleft$,$\blacktriangledown$,$\mycircle{black}$,$\mysquare{black}$) and miscible silicone oil drops ($\MyDiamond$) coalescing with oil films for (a) $Oh = [2.5-57]$ with fixed film thickness $H/R = 0.2$ and (b) $Oh \approx 11$ with $H/R = [0.2 - 4.4]$.
    Drop:film viscosity ratios are fixed at 1:10.
    (a) Increases in $Oh$ for constant $H/R$ are shown to decrease the power-law slope from $4/5$ to $2/5$, following a similar convergence trend as shown in the simulations. The decrease in velocity of $r/R$ in time for $Oh=57$ near $T/t_\eta = 0.07$ is thought be a result of the needle tip influencing the curvature of the pre-detached drop.
    (b) Increases in $H/R$ for constant $Oh$ are shown to have no effect on the power-law slope, as predicted from the simulations. Comparison with miscible oil-oil drop-film coalescence for similar viscosities as the immiscible case recovers the same power-law slope $\beta = 3/5$. Here the local curvature of the oil film near the contact point is thought to drive the spreading, limited only by viscous dissipation.
    }.
    \label{fig:experimentalData}
\end{figure}

Figures \ref{fig:experimentalData}(a,b) show the scaled apparent contact-line radius $r/R$ against the scaled time $T/t_\eta$ extracted from the experiments for varying $Oh=[2.51-57]$ and varying film thicknesses $H/R=[0.2-4.43]$, respectively. When the film thickness is fixed at $H/R=0.2$, it can be seen from Figure \ref{fig:experimentalData}(a) that increases in the Ohnesorge number $Oh$ lead to decreases in the power-law exponent from $4/5$ to $2/5$, similar to the simulation results shown previously in Figure \ref{vis}(a).
Here, the exponents are fitted from the data for $r/R$ in the range $0.4$ to $1$. 
Additionally, Figure \ref{fig:experimentalData}(b) shows the effect of increasing film thicknesses $H/R=[0.2-4]$ when $Oh \approx 11$, and we see that the evolution of the apparent contact-line radius can be described by the same power-law exponent, $r \sim t^{3/5}$, as shown by the simulations of Figure \ref{vis}(b). Here we find that varying $H/R$ has no effect on the evolution of the apparent contact line during the apparent spreading, supporting the previous conclusion that the spreading is affected only by the small region of high curvature near the 3-phase line. 

Figure \ref{fig:experimentalData}(b) also shows the results of a miscible oil drop coalescing with an oil film. Interestingly, we see that when the viscosity ratio between the drop and the film are the same as the immiscible case that the temporal evolution of the liquid bridge's radius formed during this miscible coalescence evolves with the same power-law exponent.
As was previously shown by \citet{aarts2005hydrodynamics} for viscously limited drop-film coalescence, the coalescence velocity is set by the capillary velocity $\sigma/\eta$, driven by the interfacial curvature and limited by the viscosity.
Here the temporal collapse of the interfacial dynamics for the miscible drop-film coalescence and immiscible drop-film engulfment
can be explained if the relevant capillary velocity is $\sigma_{oa}/\eta_o$, and the limiting viscosities of the film, the drop, and the air comparable between the miscible and immiscible cases. As the surface tension of the oil is at least half that of the drop in the immiscible case, this collapse shows that it is the curvature of the oil film in the neck region and subsequently the oil's surface tension that is relevant to the apparent spreading dynamics.



\section{Conclusion}
Measuring the flow field of a liquid spreading on another liquid is experimentally challenging, whereas via simulations, the internal velocity field can be easily extracted rendering trivial calculations of the viscous dissipation as well as the extraction of the drop's center of mass, the interfacial boundaries and the location of the 3-phase line. 
In this paper we utilized both simulations and experiments to investigate the effect of the film thickness and the Ohnesorge number $Oh$ on the apparent spreading and engulfment of a water drop on an oil film in both the viscously limited $Oh>1$ and inertially limited $Oh \ll 1$ regimes. 
Here we simulated the three phase flow by using the conservative phase field equation \citep{chiu2011conservative,geier2015conservative} solved by the LBM \citep{he1997theory,mohamad2011lattice}, which has recently been proposed to have improved mass conservative properties \citep{zheng2014shrinkage,geier2015conservative,aihara2019multi,baroudi2021simulation}. 

In the inertial-capillary regime, $Oh<1$, we found that post drop-film contact for film thicknesses $H/R<1$ that
the drop's center of mass would oscillate during engulfment and for film thicknesses $H/R>1$ that inertial-capillary waves on the surface of the oil film would be produced.
In the visco-capillary regime, $Oh>1$, we found that decreases in the film thickness 
led to increases in the time required for the drop to become fully engulfed. 

During the apparent spreading, we found that regardless of $Oh$ or film depth, the 3-phase line characterized by the apparent spreading radius $r$ evolved in time with a power-law scaling of $r/R\sim (T/t_\rho)^{1/2}$ in the interial-capillary regime, consistent with prior work for drops spreading on thin oil films \citep{carlson2013short}.
In the visco-capillary regime, we similarly saw that the power-law scaling during spreading was invariant to the film thickness, however variations in $Oh$ for $Oh = [1-10]$ varied the power-law evolution of the apparent contact line $r/R\sim (T/t_\eta)^{\beta}$ with $\beta$ ranging from $4/5$ for $Oh \approx 1$ to $2/5$ for $Oh>10$.
We rationalize that the invariance to film thickness on the power-law spreading of the drop's apparent contact line is a result of the viscous dissipation being concentrated locally at the contact line during the early-time spreading.

We compared our simulation results to experiments in the visco-capillary regime, validating both the qualitative shape of the drop during spreading as well as the power-law evolutions of the apparent contact-line over a similar range of $Oh$. Interestingly, we found that the coalescence of a miscible oil drop with an oil film with the same viscosity ratio as the immiscible water drop and oil film case recovered the same temporal evolution of the apparent spreading radius in time. We rationalize that this collapse of the immiscible and miscible cases are a consequence of the similar curvatures of the oil film near the apparent contact line coupled with the similar resistances to spreading by the liquid phase's viscosities.

Wetting on soft materials, such as the SLIPS surfaces considered here, is becoming increasingly relevant to the fabrication of self-healing, low-hysteresis, water-repellent surfaces. However, drop engulfment as well as many other soft-wetting phenomena are far from trivial when considered experimentally. In this work we have shown that simulations utilizing the Lattice-Boltzmann method are able to accurately capture the spreading and engulfment dynamics during soft-wetting on SLIPs surfaces, opening the door for future research probing increasingly complicated systems with various rheological properties and geometries free from the challenges of physical observation.

\section*{Declaration of Interests}
 The authors report no conflict of interest.

\section*{Acknowledgments}
We acknowledge the financial support of the Research Council of Norway through the program NANO2021 (project number 301138) and the PIRE project “Multi-scale, Multi-phase Phenomena in Complex Fluids for the Energy Industries”, founded by the Research Council of Norway and the National Science Foundation of USA under Award Number 1743794.

\appendix
\section{}
\subsection{Comparison between 2D and 3D simulations at drop engulfment dynamics}\label{app1}
\begin{figure*} 

   \includegraphics[width=\linewidth]{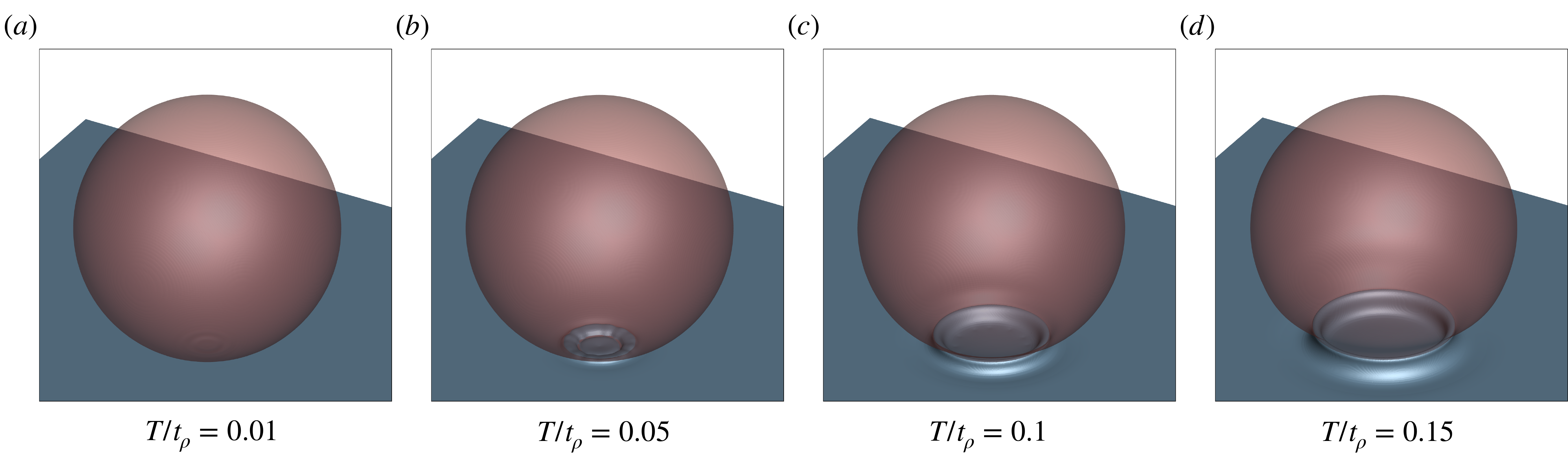}

    \caption{\label{ap1} 3D simulation of the drop dynamics (H=0.03R) at (a) $T/t_\rho=0.01$; (b) $T/t_\rho=0.05$; (c) $T/t_\rho=0.1$; (d) $T/t_\rho=0.15$, for $Oh=0.2$.}

\end{figure*}
\begin{figure*}
  \centering

  \includegraphics[width=\linewidth]{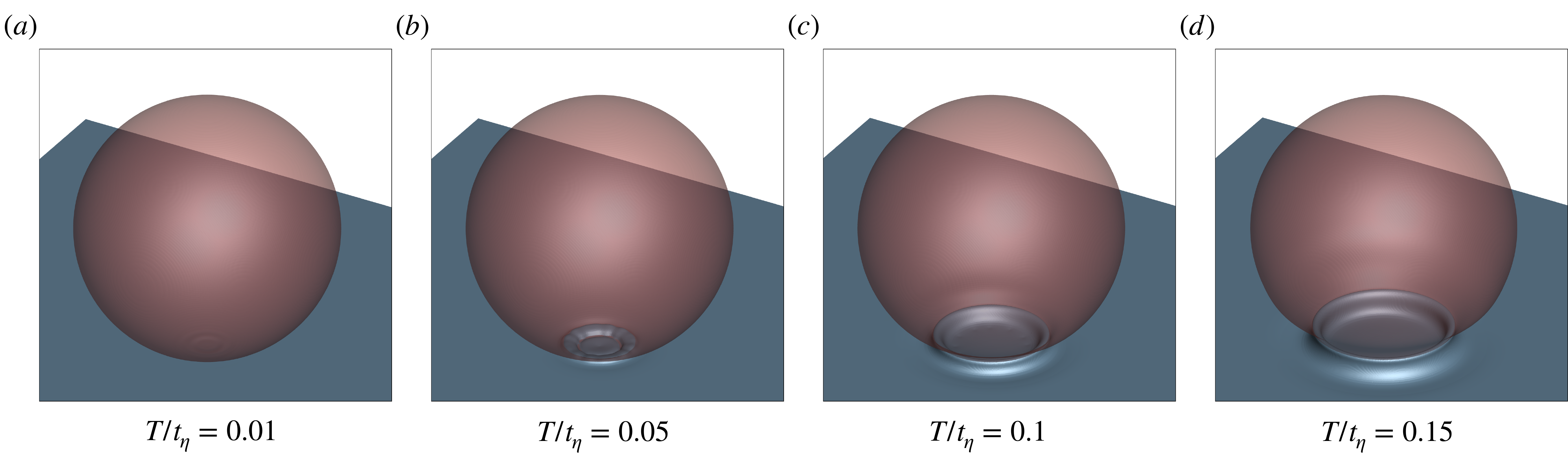}
    \caption{\label{ap2} 3D simulation of the drop dynamics (H/R=0.03) at (a) $T/t_\eta=0.01$; (b) $T/t_\eta=0.05$; (c) $T/t_\eta=0.1$; (d) $T/t_\eta=0.15$, for $Oh=4.8$.}

\end{figure*}

\begin{figure*}

  \centering
  \includegraphics[width=\linewidth]{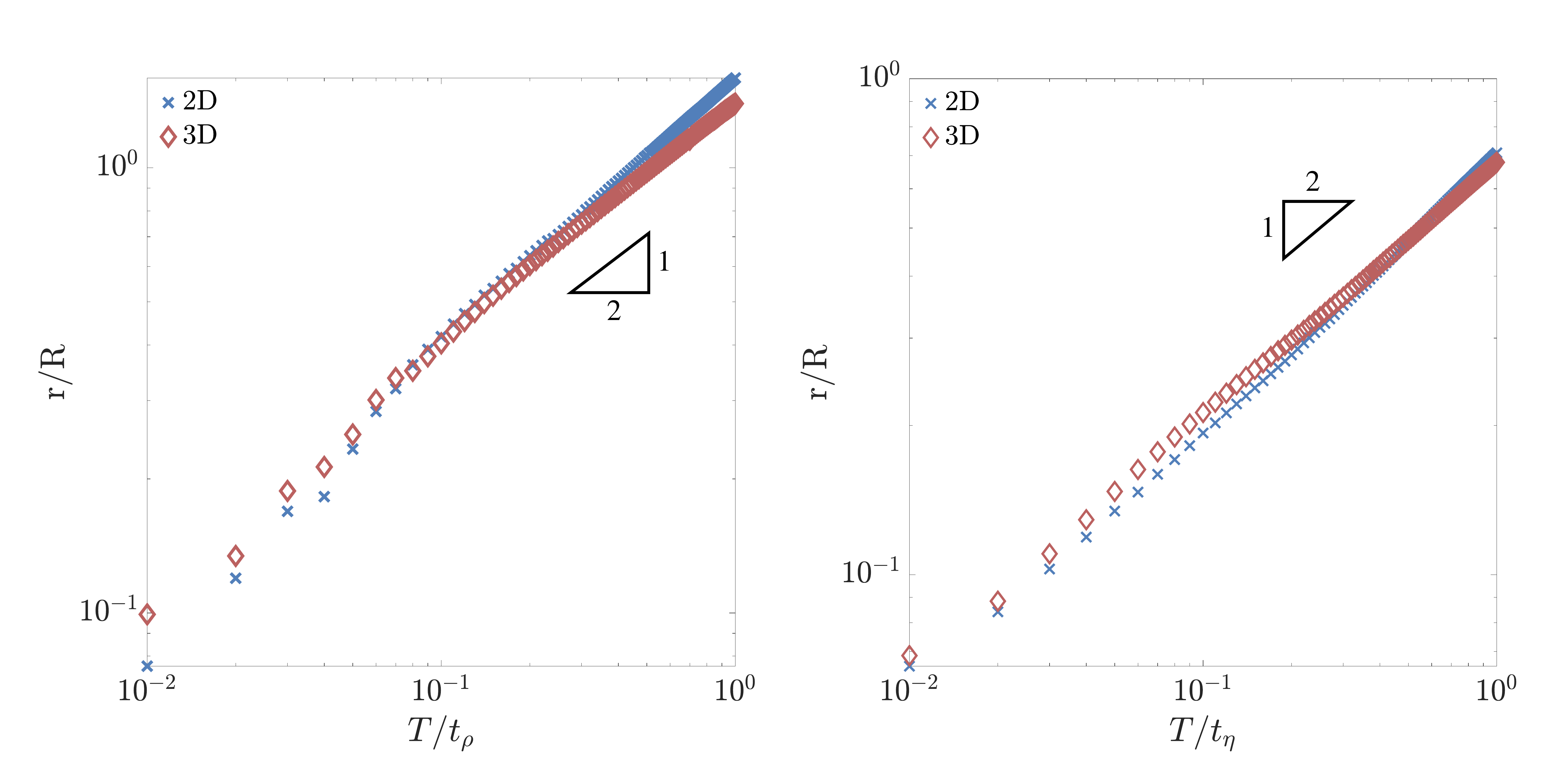}

    \caption{\label{fig10} log-log representative of the spreading radius of the water drop on the oil film with thickness $H/R=0.03$ for (a) $Oh=0.2$ and (b) $Oh=4.8$. Both simulation results indicate the power-law $r/R\sim(T/t_c)^{1/2}$.}
\end{figure*}

The short time contact line dynamics on solid or coalescence of drops have been shown to be described by 2D theory \citep{eggers1999coalescence,eddi2013short}. To verify that our 2D simulations capture well the drop dynamics for this soft wetting problem, we have conducted both 2D and 3D simulations of drops on the thin oil film which $H/R=0.03$ and compare their results. 

3D simulations of the drop evolution on the film with $H/R=0.03$ for $Oh=0.2$ and $Oh=4.8$ are shown in Figure \ref{ap1} and Figure \ref{ap2}. log-log representation of 2D and 3D drop spreading radius is shown in Figure \ref{fig10} (a) and Figure \ref{fig10} (b), which shows only a very minor effect of the 3D flow and both predict the radius to follow $r/R\sim(T/t_c)^{1/2}$.

\subsection{Materials }

Glass slides (Avantor, Article\# 631-1550P, 76 mm x 26 mm), 1000 St silicone oil VIS-RT100K (European Article\# 99515.271), 20 cSt silicone oil (CAS\# 63148-62-9) and Glycerine (CAS\# 56-81-5) were purchased from VWR International (Oslo, Norway).
30 gauge syringe tips (Weller, Mfr. Part \# KDS3012P) were purchased from RS components (Oslo, Norway).

\subsection{Chapman-Enskog Analysis}\label{C-E}

We present the Chapman-Enskog Analysis based on the Discrete Boltzmann equation in this section. The Discrete Boltzmann equation for momentum equation is given as Eq.\ref{2:2}:
\begin{equation}\label{Ap:1}
  \frac{\partial g_\alpha }{\partial t}+\boldsymbol{e}_\alpha \cdot \nabla g_\alpha=-\frac{g_\alpha-g^{eq}}{\lambda}+F_\alpha.
\end{equation}
We consider $\delta t$ as the small parameter in this case, thus the fundamental expansions based on $\delta t$ for distribution function and the time derivative are expressed as follows:
\begin{equation}\label{Ap:2}
  g_\alpha(\boldsymbol{x},t) =g_\alpha^{eq}(\boldsymbol{x},t)+\delta t g_\alpha^{(1)}(\boldsymbol{x},t)+\delta t^2g_\alpha^{(2)}(\boldsymbol{x},t),
\end{equation}
\begin{equation}\label{Ap:3}
\partial_t=\partial_{t0} +\delta t \partial_{t1}.
\end{equation}

The $\delta t$ order equation after the math calculation is then:
\begin{equation}\label{Ap:4}
  \frac{\partial g_\alpha^{eq}}{\partial t_0}=\frac{g_\alpha^{(1)}}{\tau}+F_\alpha,
\end{equation}
whereas the $\delta t^2$ order equation can be derived as:
\begin{equation}\label{Ap:5}
  \frac{\partial g_\alpha^{eq}}{\partial t_1}+(\partial t_0+\boldsymbol{e}_\alpha\cdot\nabla)g_\alpha^{(1)}=\frac{g_\alpha^{(2)}}{\tau}.
\end{equation}
The partial differential equation after the summation of $O(\delta t)+\delta t O(\delta t^2)$ can then be presented as:
\begin{equation}\label{Ap:6}
  \frac{\partial g_\alpha^{eq}}{\partial t}+\boldsymbol{e}_\alpha\cdot\nabla g_\alpha^{eq}+\delta t(\partial_{t0}+\boldsymbol{e}_\alpha\cdot\nabla)g_\alpha^{(1)}=-\frac{1}{\lambda}(g_\alpha-g_\alpha^{eq})+F_\alpha.
\end{equation}
We restrict the moments of the equilibrium distribution to obtain the macroscopic value from the distribution function:
\begin{equation}\label{Ap:7}
  \sum_{\alpha}g_\alpha^{eq}=\bar{p},
\end{equation}
\begin{equation}\label{Ap:8}
  \sum_{\alpha}g_\alpha^{eq}\boldsymbol{e}_\alpha=\boldsymbol{u}c_s^2,
\end{equation}
\begin{equation}\label{Ap:9}
 \sum_{\alpha}g_\alpha^{eq}\boldsymbol{e}_\alpha \boldsymbol{e}_\alpha=\boldsymbol{u}c_s^2+\bar{p}c_s^2,
\end{equation}
and the moments of the source term:
\begin{equation}\label{Ap:10}
  \sum_\alpha F_\alpha=-\boldsymbol{u}\cdot\nabla\bar{p},
\end{equation}
\begin{equation}\label{Ap:11}
\sum_\alpha F_\alpha \boldsymbol{e}_\alpha =\frac{c_s^2}{\rho}\left(-\nabla P+\rho\nabla\bar{p}+\nu(\nabla\boldsymbol{u}+\nabla\boldsymbol{u}^T)\nabla\rho+\boldsymbol{F}_s\right),
  \end{equation}
\begin{equation}\label{Ap:12}
\sum_\alpha F_\alpha \boldsymbol{e}_\alpha\boldsymbol{e}_\alpha =c_s^2\boldsymbol{u}\cdot\nabla\bar{p}.
  \end{equation}

Under these conditions, the following equations can be deducted from the zeroth and the first moments of Eq.\ref{Ap:6} :
\begin{equation}\label{Ap:13}
  \frac{\partial \bar{p}}{\partial t}+\nabla\cdot\boldsymbol{u}c_s^2+\boldsymbol{u}\cdot\nabla \bar{p}=0,
\end{equation}
\begin{equation}\label{Ap:14}
  \frac{\partial\boldsymbol{u}}{\partial t}+\nabla\cdot\boldsymbol{u}\boldsymbol{u}+\frac{\delta t}{c_s^2}\nabla\cdot\Pi^{(1)}=-\frac{1}{\rho}\nabla P+\frac{\nu}{\rho}(\nabla\boldsymbol{u}+\nabla\boldsymbol{u}^T)\nabla\rho+\frac{\boldsymbol{F}_s}{\rho}.
\end{equation}

We further decompose Eq.\ref{Ap:13} into continuity equation and pressure evolution equation. When we consider the kinematic viscosity $\nu=\tau c_s^2\delta t$, Eq.\ref{Ap:14} becomes:
\begin{equation}\label{Ap:15}
  \frac{\partial\boldsymbol{u}}{\partial t}+\nabla\cdot\boldsymbol{u}\boldsymbol{u}=-\frac{1}{\rho}\nabla P+\frac{1}{\rho}\nabla\cdot\eta(\nabla\boldsymbol{u}+\nabla\boldsymbol{u}^T)+\frac{\boldsymbol{F}_s}{\rho},
\end{equation}
where $\eta=\nu\rho$ is the dynamic viscosity

In the final, the governing equations, Eq.\ref{Ap:14} and Eq.\ref{Ap:15}, are retrieved from the Discrete Boltzmann equation Eq.\ref{Ap:1}.

\bibliographystyle{jfm}
\bibliography{Main.bib}

\end{document}